\documentclass[preprint]{revtex4}
\usepackage{graphicx}
\usepackage{amsmath}
\usepackage{amssymb}

\begin{document}


\title{Electrolytically Generated Nanobubbles on HOPG Surfaces}

\author{Shangjiong Yang$^{\ddag,\S}$, Peichun Tsai$^{\ddag}$, E. Stefan Kooij$^{\S}$, Andrea Prosperetti$^{\ddag,\dag}$, Harold J. W. Zandvliet$^{\S}$, and Detlef Lohse$^{\ddag}$. }

\affiliation {$^{\ddag}$Physics of Fluids Group; $^{\S}$Solid
State Physics Group, Faculty of Science and Technology, MESA+
Institute for Nanotechnology, University of Twente, 7500AE
Enschede, The Netherlands.
\\$^{\dag}$Department of Mechanical
Engineering, Johns Hopkins University, Baltimore, MD 21218, USA.}
\date{\today}

\begin{abstract}

Electrolysis of water is employed to produce surface nanobubbles
on highly orientated pyrolytic graphite (HOPG) surfaces. Hydrogen
(oxygen) nanobubbles are formed when the HOPG surface acts as
negative (positive) electrode. Coverage and volume of the
nanobubbles enhance with increasing voltage. The yield of hydrogen
nanobubbles is much larger than the yield of oxygen nanobubbles.
The growth of the individual nanobubbles during the electrolysis
process is recorded in time with the help of AFM measurements and
correlated with the total current. Both the size of the individual
nanobubbles and the total current saturate after typical 1 minute;
then the nanobubbles are in a dynamic equilibrium, meaning that
they do not further grow, in spite of ongoing gas production and
nonzero current. The surface area of nanobubbles shows a good
correlation with the nanobubble volume growth rate, suggesting
that either the electrolytic gas emerges directly at the
nanobubbles' surface, or it emerges at the electrode's surface and
then diffuses through the nanobubbles' surface. Moreover, the
experiments reveal that the time constants of the current and the
aspect ratio of nanobubbles are the same under all conditions.
Replacement of pure water by water containing a small amount of
sodium chloride ($0.01$ M) allows for larger currents, but
qualitatively gives the same results.

\end{abstract}

\maketitle

\subsection*{Introduction}

{\em Nanobubbles}, nanoscopic gas bubbles present at solid-liquid
interfaces~\cite{ball,vinogradova1,tyrrell,attard2,holmberg,simonsen,vinogradova2,agrawalpark,zhangmaeda,zhangzhangzhang,neto,zhangquinn,zhangkhan,yang1,borkent,yang2},
are in many ways fascinating objects in the field of surface
science and nanofluidics. It has been conjectured that they are
relevant for a number of phenomena and technical applications,
e.g., the liquid slippage at
walls~\cite{neto,lauga,vinogradova3,bunkin,degennes}, the
stability of colloidal systems~\cite{nguyen}, and the
nanometer-scale attractive force between hydrophobic surfaces in
solutions~\cite{vinogradova1,tyrrell,parker,attard,tyrrellattard,considine,yakubov}.
Studies on various physical aspects of nanobubbles have been
increasingly undertaken in the last few
years~\cite{tyrrell,yang1,holmberg,borkent,simonsen,vinogradova2,yang2,agrawalpark,zhangzhangzhang,zhangmaeda,ishida,zhangzhanglou,agrawalmckinley,switkes,steitz}.
The solid surfaces employed in these studies include
gold~\cite{holmberg},
polystyrene~\cite{simonsen,vinogradova2,agrawalpark},
mica~\cite{zhangzhanglou}, silane-hydrophobilized silicon
wafer~\cite{zhangmaeda,agrawalmckinley,yang1,switkes}, and highly
orientated pyrolytic graphite
(HOPG)~\cite{zhangzhangzhang,zhangmaeda,yang2}. Most studies are
done with highly purified water (Milli-Q), though some experiments
have been done with alcohol~\cite{simonsen} or dilute sulfuric
acid solution~\cite{zhangzhangzhang}. Atomic force microcopy (AFM)
in tapping mode is adopted in most
experiments~\cite{tyrrell,yang1,holmberg,borkent,simonsen,yang2,agrawalpark,zhangzhangzhang,zhangmaeda,zhangzhanglou,agrawalmckinley,ishida},
while other techniques such as rapid cryofixation-freeze
fracture~\cite{switkes}, neutron reflectometry~\cite{steitz},
high-energy x-ray reflectivity~\cite{mezger}, and internal
reflection infrared spectroscopy~\cite{zhangquinn} have also been
employed. Experimental observations show that nanobubbles are very
stable, having an extraordinary shape with remarkably large aspect
ratio~\cite{zhangmaeda,yang1} which even further increases with
decreasing nanobubble size~\cite{borkent2}. The lifetime of
nanobubbles shows a dependence on the gas type~\cite{zhangquinn}.
Besides the surface hydrophobicity, the spatial dimensions of the
hydrophobic domains on the surface are crucial for the formation
of nanobubbles~\cite{agrawalmckinley}. It has also been reported
that the formation of nanobubbles is related to surface
nanostructures: the majority of nanobubbles prefer to form in the
vicinity of nanometer-deep grooves~\cite{yang1} or on the upper
side of atomic steps~\cite{yang2} on the surfaces. In addition, an
increase of substrate temperature, water temperature, or gas
concentration in water increases the density and size of
nanobubbles~\cite{yang1,zhangzhanglou}. These observations clearly
reveal that the formation of nanobubbles is very sensitive to
surface and liquid conditions. Yet, is there a simple method that
leads to the controlled formation and growth of nanobubbles?

In electrochemical reactions, gas molecules are generated at
electrode surfaces. Most studies have hitherto focused on mini- or
micrometer sized bubbles, which are formed at and subsequently
detach from the electrodes;
see~\cite{tsai,boissonneau,fan,volanschi,gabriellimacias,gabrielli}
and references therein. The interest originates partly from the
significant influence of the bubbles on reaction systems. E.g.,
convection caused by the evolution of electrogenerated
microbubbles increases electrolyte flow and can enhance production
processes~\cite{boissonneau}. The interest in electrochemically
generated nanobubbles is more recent. It has been hypothesized
that the existence of nanobubbles at electrode surfaces favors the
formation of submicrometer-sized vaterite tubes in the
electrolysis-induced mineralization~\cite{fan}. Zhang et
al~\cite{zhangzhangzhang} confirmed that electrochemical
generation of hydrogen induces the formation of nanobubbles on the
electrode surface in sulfuric acid solution.

The work described in this article is motivated by two issues: (i)
Electrolysis of water is a reliable and controllable way to
rapidly produce high local gas concentration at the electrode
surfaces. Gas concentration significantly affects the formation of
nanobubbles~\cite{yang1}. Electrolysis of water therefore is an
easy method to control the appearance and growth of surface
nanobubbles. This is demonstrated in this article by performing
AFM measurements of nanobubbles on an HOPG surface which acts as
electrode. To reduce the effect of any possible impurities in the
liquid, since nanobubbles are extremely sensitive to surfactants,
ultraclean water (see below for qualification) is used as
electrolyte. In addition, to test the reproducibility, an aqueous
sodium chloride solution ($0.01$ M) is also used. We study the
bubble coverage, volume, size, and aspect ratio at different
voltages. In addition, we show the real-time development of
individual nanobubbles, before they finally achieve a dynamic
equilibrium condition. Remarkably, the nanobubble's surface area
and its volume growth are highly correlated, suggesting that
either the electrolytic gas is produced at the whole surface of
the nanobubbles, or it is generated at the electrode's surface and
diffuses to the surface of nanobubbles. (ii) The second issue of
this article is to correlate geometric feature of the nanobubbles
with the electric current that flows between the two electrodes.
We find a good correlation between the aspect ratio of the
nanobubbles and the current.

\subsection*{Experimental Section}

The water is prepared by a Milli-Q Synthesis A10 system (Millipore
SAS, France) and then degassed at 1 mbar for 4 hours. AFM
measurements are done with a PicoSPM (Molecular Imaging, AZ USA)
operated in tapping mode. Excitation of the tip vibration is done
acoustically, using a small piezo-element in the tip holder. The
AFM operating parameters in water are as follows: scanning speed
$6\,\mu{\rm m/s}$; free amplitude 400 mV; set-point amplitude 300
mV; resonance frequency 20 kHz. AFM scanning is performed by a
hydrophilic ${\rm Si}_3{\rm N}_4$ ultra-sharp AFM tip (radius of
curvature $<$ $10\,{\rm nm}$, full tip cone angle $<$ $30^\circ$,
NSC18/AlBS, MikroMasch, France; rinsed with ethanol and pure water
before use). An HOPG sample (HOPG ZYB/$1.75$, size $10\,{\rm mm}
\times 10\,{\rm mm}$, MikroMasch, France) with a freshly cleaved
surface placed on a copper plate is used as nanobubble forming
surface and at the same time as one of the electrodes.  A platinum
wire (diameter $0.25$ mm) placed next to the AFM cantilever is
used as the other electrode. The copper plate and the platinum
wire are connected to an electrometer (Picoammeter/Voltage Source
6478, Keithley Instruments Inc., OH USA). After a water drop
(volume $0.33$ ml - $0.40$ ml) is placed on the HOPG surface and
the desired voltage is imposed, AFM scanning process is started
immediately. Figure 1 shows a sketch of the setup. When the HOPG
sample acts as the negative electrode (cathode), the reduction
process of water leads to the formation of hydrogen molecules on
the HOPG surface, $2H_2O(l)+2e^- \rightarrow H_2(g)+2OH^-(aq)$.
Oxygen molecules are produced on the HOPG surface when the HOPG
sample is switched to be the positive electrode (anode) and
therefore the oxidation process of water on the surface leads to
oxygen molecules, $2H_2O(l) \rightarrow O_2(g)+4e^-+4H^+(aq)$. The
experiments are carried out in a standard lab environment with a
temperature between 20 and $23$ $^\circ$C. The temperature change
of the HOPG sample during the measurements is less than $0.1$ K.

\subsection*{Results and Discussion}

\subsubsection*{Nanobubbles by Electrolysis of Water: Dependence on
Applied Voltage and Gas Type}

Previous experimental results show that no nanobubbles are formed
on HOPG surfaces unless the so-called ethanol-water-exchange step
is carried out~\cite{zhangmaeda,yang2}. This is due to the
hydrophilic nature of the surface (macroscopic contact angle $<
90^\circ$) that disfavors the attachment of surface bubbles.
Electrolysis of water can be a robust method for a sufficient
yield of nanobubbles on HOPG~\cite{zhangzhangzhang}. AFM
measurements by tapping mode are performed on the HOPG surface.

Figure 2 shows the topography images of the HOPG surface used as
cathode with different applied voltages: (a) $1\,{\rm V}$, (b)
$1.5\,{\rm V}$, (c) $2\,{\rm V}$, (d) $2.5\,{\rm V}$, (e) $0\,{\rm
V}$, (f) $3\,{\rm V}$, (g) $3.5\,{\rm V}$, (h) $4\,{\rm V}$; the
height range for each image is (a-b) $42\,{\rm nm}$, (c-e)
$50.6\,{\rm nm}$, (f) $61.2\,{\rm nm}$, and (g-h) $115.5\,{\rm
nm}$. The images are recorded continuously from (a) to (h), with a
scanning time of $8.5\,{\rm min}$ per image. Nanobubbles
(hydrogen) form with varying density at different voltages. The
atomic steps of HOPG are visible when the nanobubble coverage is
low and thus act as a good reference position at the nanoscale
when conducting AFM measurements. The formation of nanobubbles
increases tremendously when increasing the voltage from $1.5$ to
$2\,{\rm V}$. Figure 2 (e) reveals that the nanobubbles remain
stable even when the voltage has been switched off from $2.5\,{\rm
V}$ (d). This demonstrates the robust stability of nanobubbles,
which is similar to the previous finding that the
heating-water-generated nanobubbles do not disappear when the
water is cooled down~\cite{yang1}. At the higher voltages,
nanobubbles cover the entire surface with much larger individual
sizes, see Figure 2 (f-h). Growth and detachment of nanobubbles
are observed under the higher electric potentials; the examples
are marked by arrows in the images. The dependence of nanobubble
coverage and volume on the applied voltage is respectively
depicted in Figure 2 (i) and (j) (error bar $\pm$ 5\%). With
increasing voltage, more hydrogen molecules are produced at the
cathode (HOPG surface), enhancing the local gas concentration.
This leads to an increase in the coverage and volume of the
nanobubbles, as revealed by plot (i) and (j) in Figure 2. At high
voltages, i.e., $4.5$ or 5 V, microbubbles developing at the HOPG
surface can already be monitored by an optical camera. The
evolution of these microbubbles ruins the AFM scanning process by
interfering with the vibrating AFM cantilever.

AFM topography images of HOPG surface used as anode are shown in
Figure 3. Different voltages are applied: (a) $1\,{\rm V}$, (b)
$1.5\,{\rm V}$, (c) $2\,{\rm V}$, (d) $2.5\,{\rm V}$, (e) $3\,{\rm
V}$, (f) $3.5\,{\rm V}$, (g) $4\,{\rm V}$ (height range: (a)
$12\,{\rm nm}$, (b-g) $35\,{\rm nm}$). The scanning time of each
image is $8.5\,{\rm min}$. The images are taken in succession from
(a) to (g) without any delay. Oxygen nanobubbles form on the
surface. Comparing to the hydrogen case in Figure 2, the
production of nanobubbles in Figure 3 is much smaller. We suggest
that this is due to the considerable difference in the solubility
of oxygen and hydrogen in water (at 20 $^\circ$C the solubility of
oxygen is $\sim2$ times higher than that of hydrogen), as well as
to the different production rate during the electrolysis, ${\rm
H}_2:{\rm O}_2=2:1$. The nanobubble coverage and volume are
plotted as functions of the applied voltage, respectively, in
Figure 3 (h) and (i) (error bar $\pm$ 5\%). For both hydrogen and
oxygen, the plots in Figure 2 (i-j) and 3 (h-i) reveal a threshold
and saturation of the nanobubble formation in dependence of the
applied voltage.

The coverage and volume values presented in Figure 2 (i-j) and 3
(h-i) are calculated by setting an appropriate hight threshold $z$
to extract nanobubbles. This is illustrated by the example in
Figure 4. AFM (tapping mode) topography images (height range
$27.2\,{\rm nm}$) of hydrogen surface nanobubbles are shown with
different thresholds $z$ applied for the identification of surface
nanobubbles: (a) $z$=0 nm, (b) $z$=6 nm, (c) $z$=7 nm, (d) $z$=8
nm, (e) $z$=9 nm, (f) $z$=10 nm, and (g) $z$=14 nm. The principle
is sketched in Figure 4 (h). Areas below the threshold are
represented as blue, whereas areas above are shown as yellowish
depending on the height. The fraction of the latter area is shown
in Figure 4 (i) as function of the threshold $z$. The curve shows
a pronounced shape. We take the value at the end of the straight
shape region (marked by an arrow), where $z$=9 nm presenting a
nanobubble identification as shown in image (e), as estimate for
the nanobubble coverage and volume statistics.

\subsubsection*{Nanobubbles in Dynamic Equilibrium}

During the experiments, each chosen voltage is continuously
applied while performing the AFM measurements shown in Figure 2
and 3. The constant voltage results in continuous charge flux
through the system. Under such a condition, one may expect that
surface nanobubbles would constantly accumulate on the electrode
surfaces. However, our AFM images (Figure 2 and 3), taken after a
certain transient time, show stationary nanobubbles of certain
sizes. In other words, electrolytically generated nanobubbles
experience a saturation in their development.

This suggests that the nanobubbles are in a {\em dynamic
equilibrium} state. There are gas-flows into and out of the
nanobubbles simultaneously, which balance each other allowing for
a constant volume. When the inflow overwhelms the outflow,
nanobubbles start to grow. This happens when the voltage is
increased, producing more charges and leading to a larger gas flow
into the nanobubbles, thus breaking the previous balance between
the inflow and the outflow, and consequently causing the
nanobubbles to grow. As the nanobubbles grow, the outflow starts
to increase till it reaches a new equilibrium state with the
inflow. The nanobubbles then again remain in a stable condition.

In order to quantify the growing process of the nanobubbles, we
focus on a number of individual nanobubbles and measure the
evolution of various geometric properties such as width, height,
aspect ratio, etc. In addition and in parallel, we measure the
global current as function of time (shown in the following
sections). The electric current decays as the nanobubbles grow.
This decrease in current, which reduces the amount of gas produced
on the surface, effectively decreases the inflow to the
nanobubbles. This of course helps to reach a new dynamic
equilibrium state, but we stress again that the current is nonzero
in the saturated state. The data of the current as function of
time and the nanobubble development show saturation on the same
time scale. At the saturated state, the nanobubble growth
terminates, whereas the saturated current is nonzero. This
observation clearly suggests the existence of a {\em dynamic
equilibrium} of the nanobubbles.

\subsubsection*{Time Evolution of Nanobubbles}

The appearance of nanobubbles can easily be controlled by an
increase of the voltage, as revealed in Figure 2 and 3. Thus, we
can capture the dynamics of the nanobubble growth by operating the
AFM tip to repeatedly scan along a fixed straight line on the
surface over the time of the electrolysis. With this method we
perspicuously quantify the evolution of the nanobubbles at the
moment of increasing voltage. The measurements are shown in Figure
5.

During the experiment, we first start the AFM scan over one line
on the HOPG surface, and then we apply the desired voltage to
generate surface nanobubbles - meanwhile the AFM scan is
continuously running. The time when we apply the voltage is taken
as 0. Each AFM line-scan takes 1 sec; the profile of the
developing nanobubble is continuously recorded. Figure 5 (a)
presents the profiles of a nanobubble generated with 1 V and the
adjacent substrate surface (HOPG, as cathode) at different time
with interval of 10 sec. Plot (b) exhibits the dynamics of another
nanobubble generated at 2 V. It is clearly shown that the
nanobubbles start to grow continuously immediately after their
appearance on the surface; this is also demonstrated by the
nanobubble area $vs.$ time plots in Figure 6 and 7. Note that the
growth terminates after 70 sec for plot (a) and after 40 sec for
plot (b) in Figure 5. The nanobubbles then remain stable, although
the voltage is still applied and the current is nonzero. The
stabilized nanobubble in Figure 5 (a) is approximately 200 nm in
width and 5 nm in height. Interestingly, the measurements show
that the nanobubbles grow with a faster rate in height than in
width. The good agreement in the topography among the profiles of
the adjacent HOPG surface at different times reveals that the AFM
measuring is not considerably perturbed by the electrolysis
process or the emergence of the nanobubbles, the profiles of the
nanobubble therefore can be compared.

From the nanobubble profiles recorded by the AFM scan, we extract
the width and height values of the nanobubbles at different times.
Note that the AFM scan does not necessarily cross the center (the
maximum width and height) of each nanobubble. Therefore the
extracted width and height values may be lower than the maximum
values. By assuming the shape of nanobubbles as a spherical cap,
we estimate the surface area of a nanobubble as $\pi$$w^2/4$ using
the extracted width $w$. In a corresponding way we estimate the
volume growth rate of a nanobubble, which as well as the surface
area is then plotted as a function of time, shown in Figure 6 (a)
and (b), respectively. Exponential fits (red lines) are applied to
both plots and values of the time constant $\tau$ are extracted.
$\tau$ values of the area and volume growth rate are plotted
versus voltage in Figure 6 (c). The plot shows that nanobubble
area and volume growth rate have a good correlation at all four
voltages. This result suggests that the electrolytically generated
gas is produced on the whole surface of the nanobubbles, implying
that the whole surface of the nanobubbles is electrically charged.
Alternatively, the electrolytically generated gas could be
produced on the electrode surface (HOPG) and subsequently diffuse
through the surface of the nanobubbles.

\subsubsection*{Correlation between Global Current and Local
Nanobubble Growth}

The global current of the electrolysis system is recorded as a
function of time with a sampling rate $0.367$ sec and an
integration time $0.102$ sec. To test the reproducibility, two
HOPG samples and three freshly cleaved surfaces on each sample are
analyzed (as cathode). Thus current measurements are done on six
different HOPG surfaces at each voltage. All these results show
that the current vs. time curves present an exponential decay at
voltages below 3 V. At higher voltages, the current fluctuates
strongly. The reason is that more and bigger bubbles are formed at
higher voltages. Growth and detachment of the bubbles cause the
current to fluctuate. This is in the agreement with the
observations in Figure 2 and
refs.~\cite{gabriellimacias,gabrielli}.

As described in the previous section, we extract the width and
height values of nanobubbles at different times, based on the
AFM-recorded profiles of the nanobubbles. We here estimate the
nanobubble area and aspect ratio (width over height), which are
then plotted as a function of time. In Figure 7, graphs show the
dynamics of current, nanobubble area, and nanobubble aspect ratio
within the first 60 sec at (a) 1 V, (b) $1.5$ V, (c) 2 V, and (d)
$2.5$ V. These three quantities are recorded simultaneously at
each voltage. The nanobubble development and the current decay are
strongly correlated. In Figure 7 (a), as an example, the
nanobubble expands rapidly in the first 20 sec, from 20 to 50 sec
it grows less quickly, thereafter it reaches a stable state, as
revealed by the area vs. time plot (red square); the current decay
behaves in a correlated way on the same timescale (black dot).
Interestingly, along with the current decay, the nanobubble aspect
ratio (green triangle) decreases too. This indicates that
nanobubbles occur initially in an ultrathin-film form with a large
aspect ratio, and then accumulate with a higher growth rate in
vertical as compared to horizontal direction. This is consistent
with the observation in Figure 5.

The gas produced at the electrode surface depends on the electric
charge passing from one electrode to the other. Figure 7 shows
that the global current reaches an equilibrium state as soon as
the nanobubble development terminates. The amount of the excess
electric charge above the equilibrium state within time (60 sec)
is estimated for voltage 1, $1.5$, 2, and $2.5$ V, respectively.
The amount is plotted against the nanobubble coverage and volume
at each voltage, as shown in Figure 8 (a) and (b), respectively.
The red lines are linear fits. Note that the fits are a guide to
eyes, not necessarily suggesting that both coverage and volume of
the nanobubbles have a linear relation with the charge. One can
see that the amount of nanobubbles produced increases as the
amount of excess electric charge increases, showing the
contribution of the gas yielded by electrolysis to the nanobubble
formation. Note the offset of the fits: in spite of the nonzero
charge there is no nanobubble production (zero nanobubble coverage
and volume). The offset indicates that part of the
electrolytically generated gas dissolves, not contributing to the
formation of nanobubbles. This crucial charge may also be needed
to build up a dielectric layer at the electrode. Zhang $et$ $al.$
reported similar observation that a formation time for nanobubbles
is required and it decreases when the applied voltage
increases~\cite{zhangzhangzhang}.

For further analysis of the time scales of the current and the
nanobubble growth, the current, the nanobubble area, and the
aspect ratio plots are fitted with an exponential. Examples are
shown in Figure 9 (a-c). Red curves are the fits to the data (blue
dots), from which the time constants $\tau$ are extracted. The
values of $\tau$ are presented as a function of voltage for area
(error $\pm$ 17\%) and current (error $\pm$ 13\%) in Figure 9 (d),
and for aspect ratio (error $\pm$ 16\%) and current in Figure 9
(e). First, we note that the $\tau$'s decrease with increasing
voltage, indicating that the development of nanobubbles and the
decay of current take place more rapidly at higher voltage. One
can moreover see that the $\tau$ values of area and current: (i)
agree well at $2$ and $2.5$ V when the nanobubble coverage is
high; hence the nanobubble growth leads to a decrease of the
current in the system; (ii) deviate at 1 and $1.5$ V, when the
nanobubble coverage is rather low. We stress that the current is a
global parameter, whereas the area of individual nanobubbles is a
local parameter. Interestingly, the nanobubble aspect ratio and
the current are perfectly correlated, as shown in Figures 7 and 9
(e). We do not have an explanation for this finding. We note that
the aspect ratio presumably exhibits a universal way of nanobubble
development. Therefore it might be a global feature.

\subsubsection*{NaCl Solution as Electrolyte}

To study the robustness of our observations, in addition to pure
water an aqueous sodium chloride (NaCl) solution ($0.01$ M) was
used as electrolyte. Using the same experimental setup as
described in Figure 1, the NaCl solution is deposited on the HOPG
surface acting as the negative electrode (cathode). With no
applied voltage, no nanobubbles are formed. When the voltage is
imposed, the formation of hydrogen nanobubbles starts to become
observable. The voltage is varied as $0.25$, $0.5$, $0.75$, $1$
and $1.25$ V. A small amount of nanobubbles are already formed at
$0.25$ V. The nanobubble formation increases tremendously as the
voltage is switched from $0.5$ to $0.75$ V. This is similar to the
result shown in Figure 2 where the formation of nanobubbles jumps
from $1.5$ to 2 V. The formation of nanobubbles in NaCl solution
starts to show a saturation after $0.75$ V. The dependence of
nanobubble coverage and volume upon the applied voltage is
depicted in Figure 10 (a) and (b) (error bar $\pm$ 5\%),
respectively. When the voltage is higher than $1.25$ V, AFM
imaging is disturbed by bigger bubbles developing or detaching
from the surface. The formation of nanobubbles in the NaCl
solution is similar to that in pure water, except that because of
the reduced resistance due to the dissolved salt, the effective
voltage is reduced by a factor of about 3: 2 V for the pure water
and $0. 75$ V for the NaCl solution. Note that the volume and
coverage of nanobubbles at the effective voltages in the two cases
are comparable.

The time evolution of nanobubbles at $0.25$ V in the NaCl solution
is shown in Figure 11. The nanobubbles continuously develop on the
surface till 40 sec and then remain stable, as revealed in Figure
11 (a). As in the experiment shown in Figure 7, the global current
of the electrolysis system, the nanobubble surface area, and the
aspect ratio are measured simultaneously as a function of time
within the first 60 sec, as shown in Figure 11 (b). A good
correlation between the current decay and the nanobubble
development is found - this is the same observation as with pure
water. The aspect ratio also shows a comparable correlation with
the current.  The experiments with the NaCl solution reproduce our
findings concerning the nanobubbles in dynamic equilibrium. Again,
good correlations between global current decay and bubble growth
dynamics are found.

\subsection*{Conclusion}

We have shown that the electrolysis of water is a reliable method
to produce both hydrogen (at cathode) and oxygen (at anode)
surface nanobubbles. Coverage and volume of the nanobubbles grow
substantially with increasing voltage. The yield of hydrogen
nanobubbles is much higher than that of oxygen nanobubbles. Our
results of nanobubble evolution have shown that nanobubbles occur
initially in an ultrathin-film with a large aspect ratio, and
subsequently grow with a higher rate in vertical rather than in
horizontal direction. In spite of the continuously applied voltage
and a nonzero current, the growth of the nanobubbles terminates
after a typical time, showing that electrolytically generated
nanobubbles are in a dynamic equilibrium condition. We note that
also the spontaneously forming nanobubbles (i.e., without
electrolysis) might be in a dynamics equilibrium, in which the gas
outflux through the Laplace pressure is compensated by a gas
influx at the contact line, as has recently been speculated in
reference~\cite{brenner}. In addition, we have found a correlation
between the surface area and the volume growth rate of
nanobubbles, suggesting possible ways how electrolytic gas emerges
on the surface. The global current as function of time is strongly
correlated with the bubble aspect ratio. The experiments with an
aqueous sodium chloride solution ($0.01$ M) give similar results.

\acknowledgments The authors thank Bram Borkent, Michael Brenner,
Bene Poelsema, and Jacco Snoeijer for stimulating discussions.
This work is part of the research program of the Stichting voor
Fundamenteel Onderzoek der Materie (FOM), financially supported by
the Nederlandse Organisatie voor Wetenschappelijk Onderzoek (NWO).

\clearpage

\begin{figure*}
\includegraphics[height=50mm]{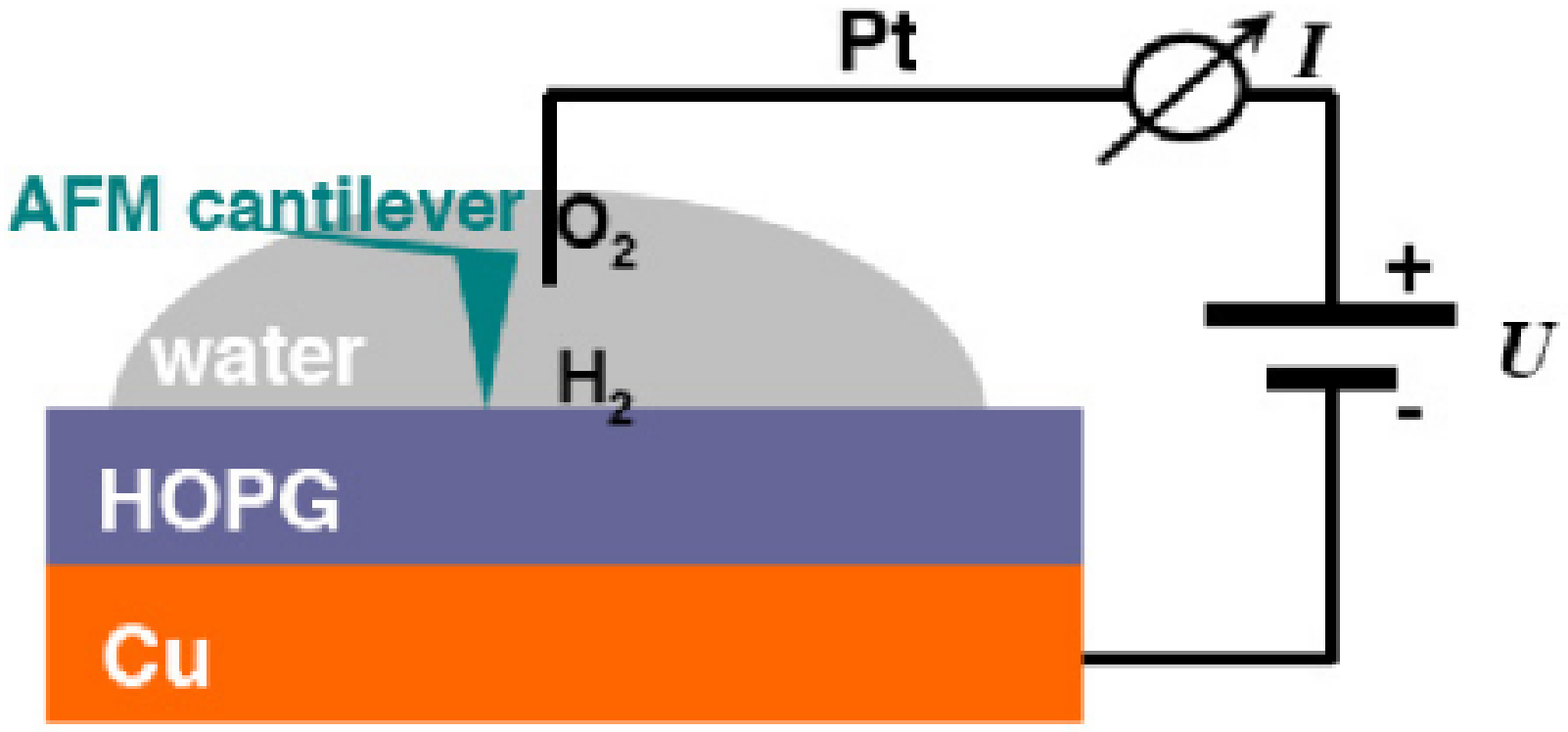}
\vspace{0mm}\\
\caption{ \label{1} (color) Sketch describing our experimental
setup. The HOPG sample is placed on a copper plate. A platinum
wire of diameter $0.25$ mm is set ($\sim 2$ mm away) next to the
AFM cantilever. The copper plate and the platinum wire are
connected to a power source supplying the voltage $U$ (the
electrometer). The platinum wire and the HOPG surface act as the
electrodes. The current $I$ is measured with a high precision
amperemeter. When the HOPG surface is used as the negative
(positive) electrode, water reduction (oxidation) process takes
place producing hydrogen (oxygen) molecules on the HOPG surface.}
\end{figure*}

\begin{figure*}
\includegraphics[height=40mm]{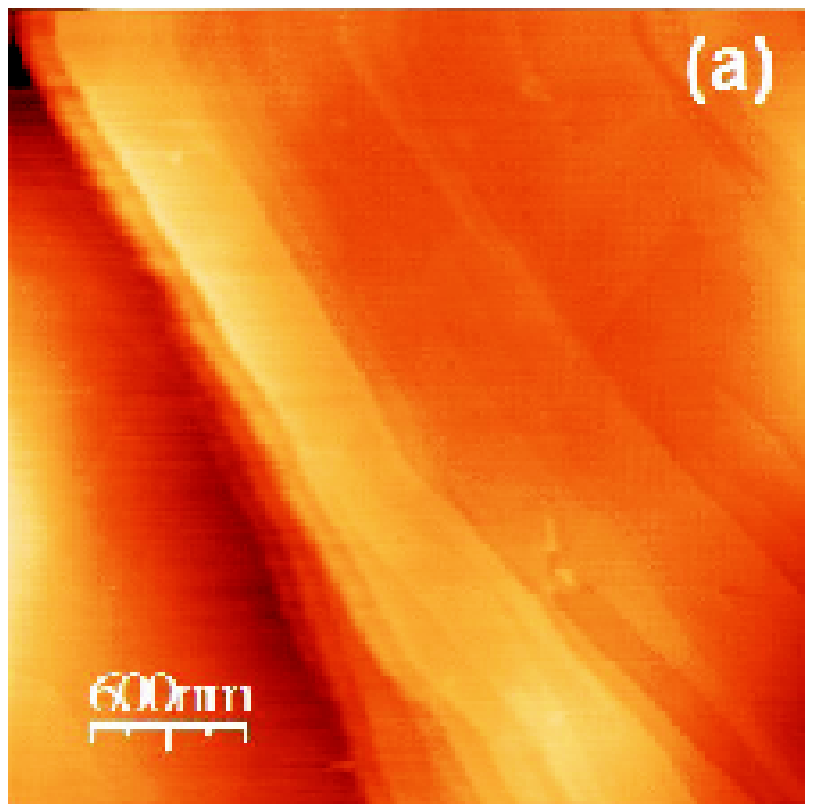}
\includegraphics[height=40mm]{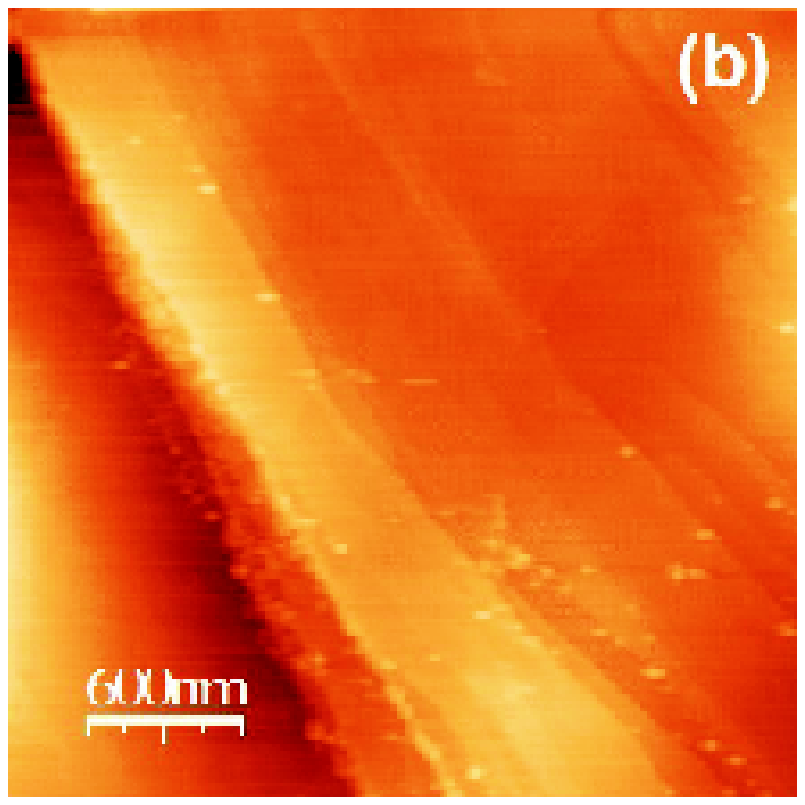}
\includegraphics[height=40mm]{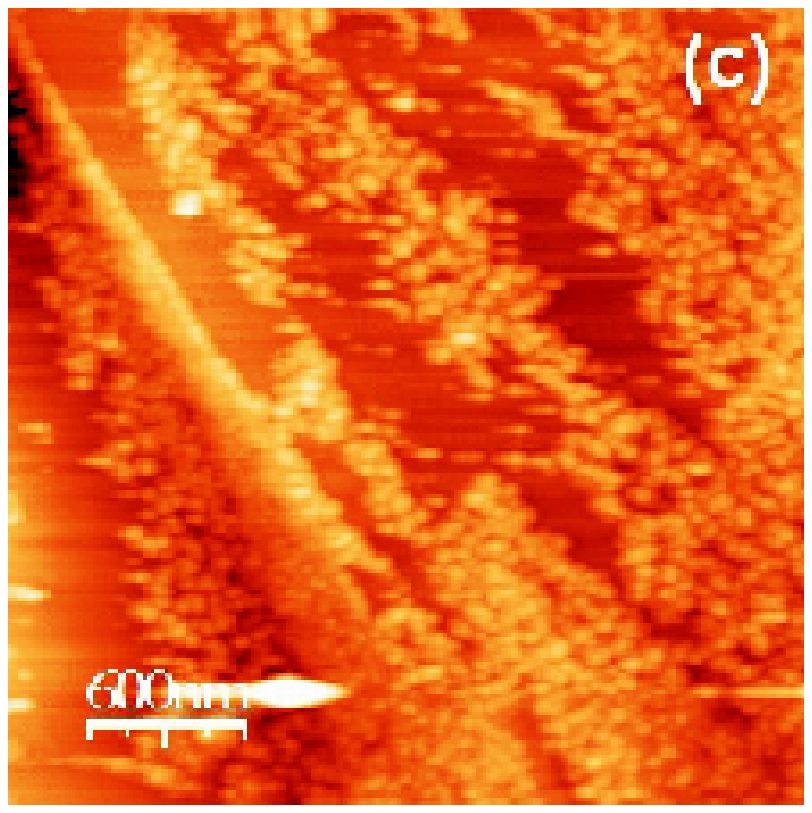}
\includegraphics[height=40mm]{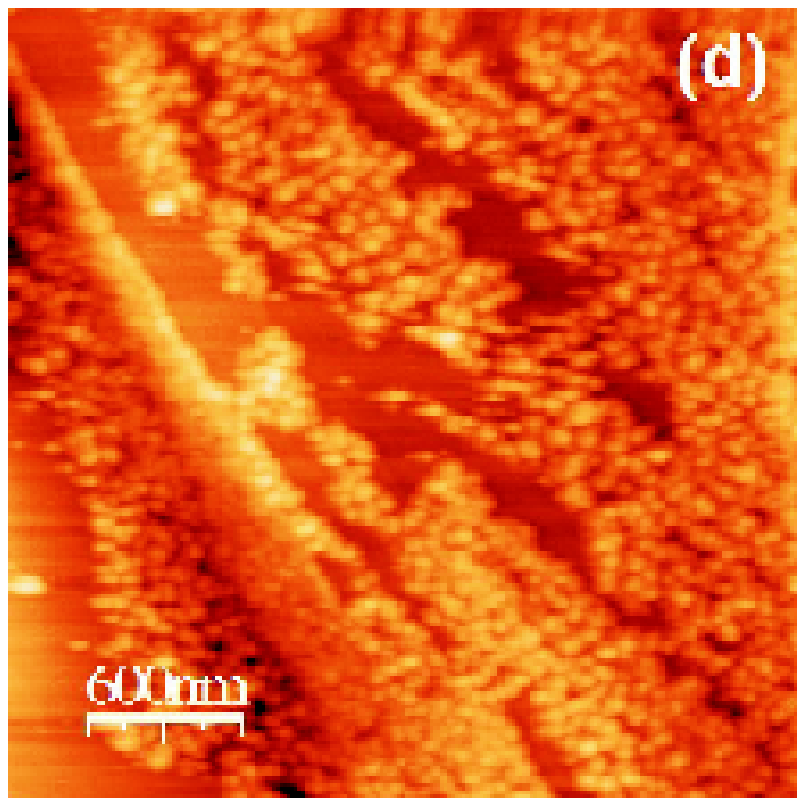}
\includegraphics[height=40mm]{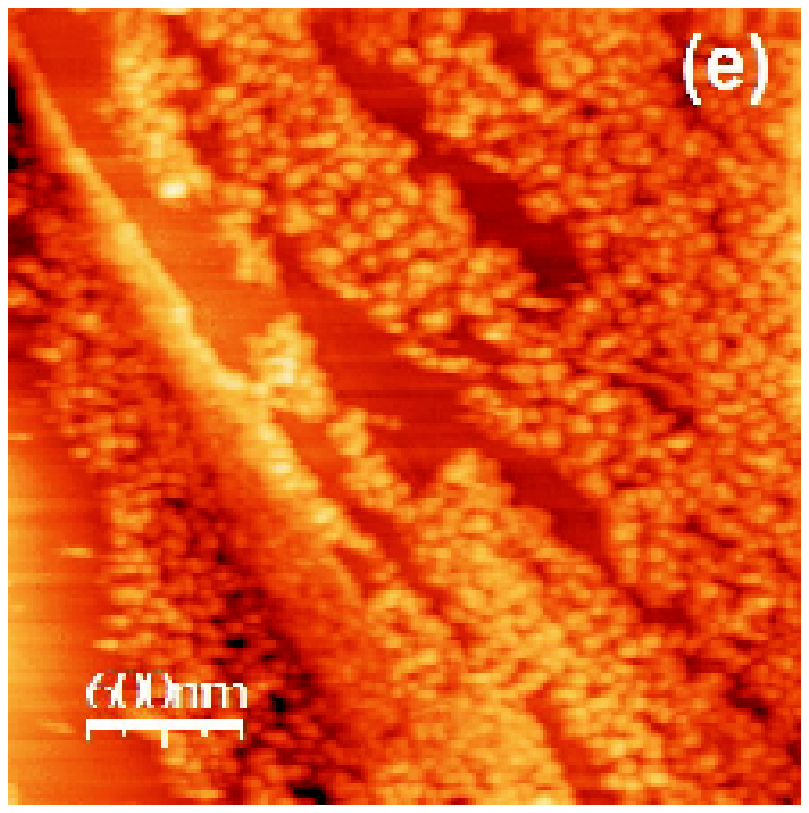}
\includegraphics[height=40mm]{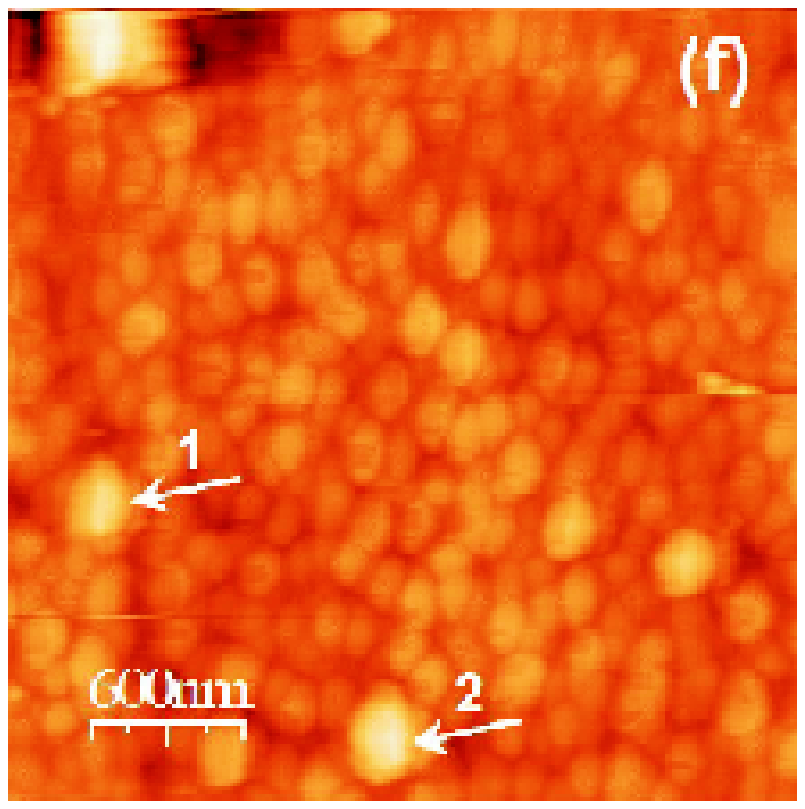}
\includegraphics[height=40mm]{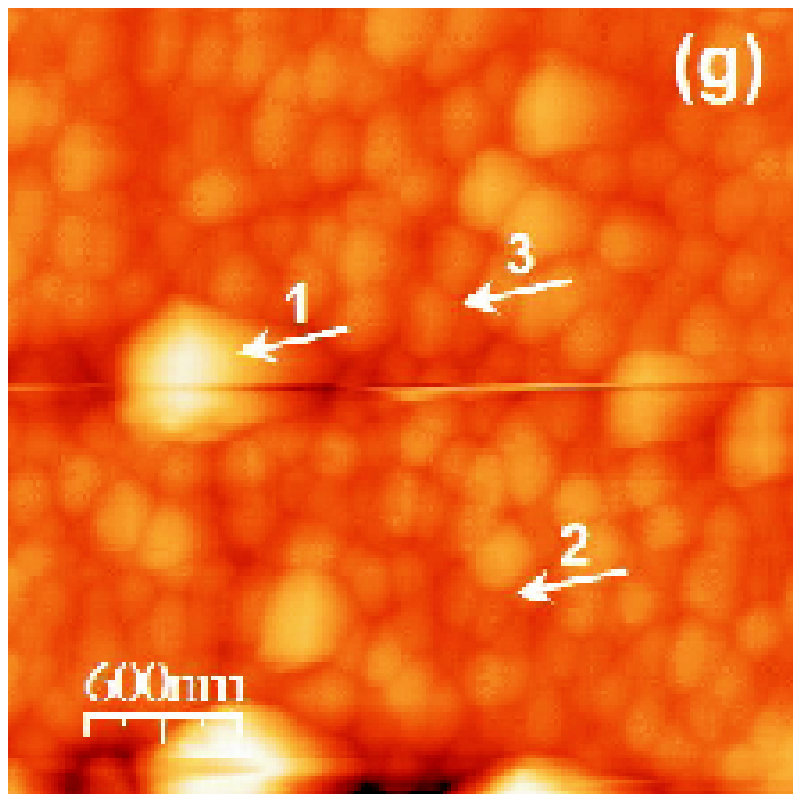}
\includegraphics[height=40mm]{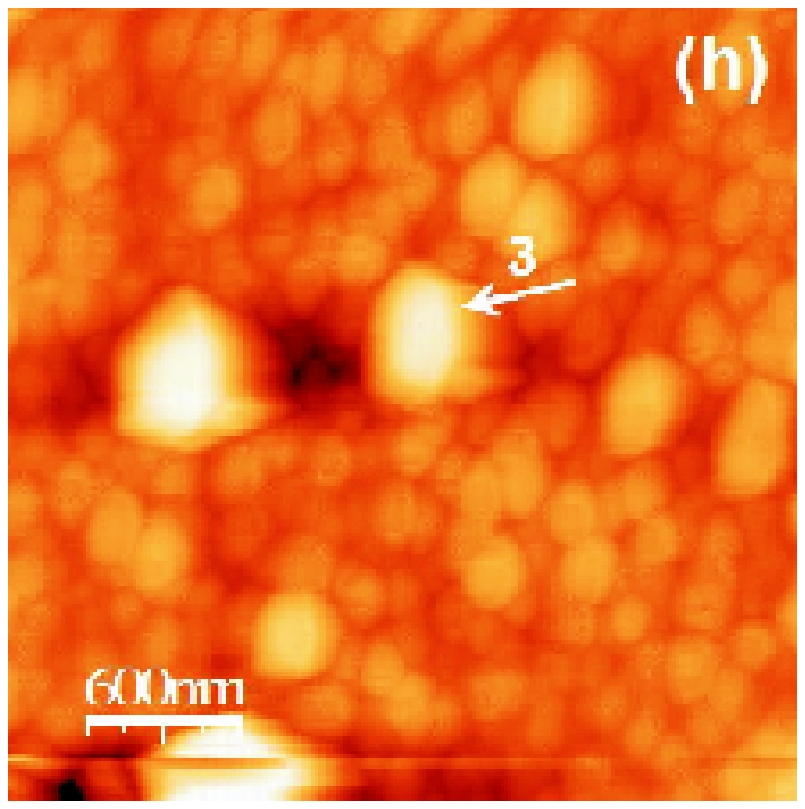}
\includegraphics[height=55mm]{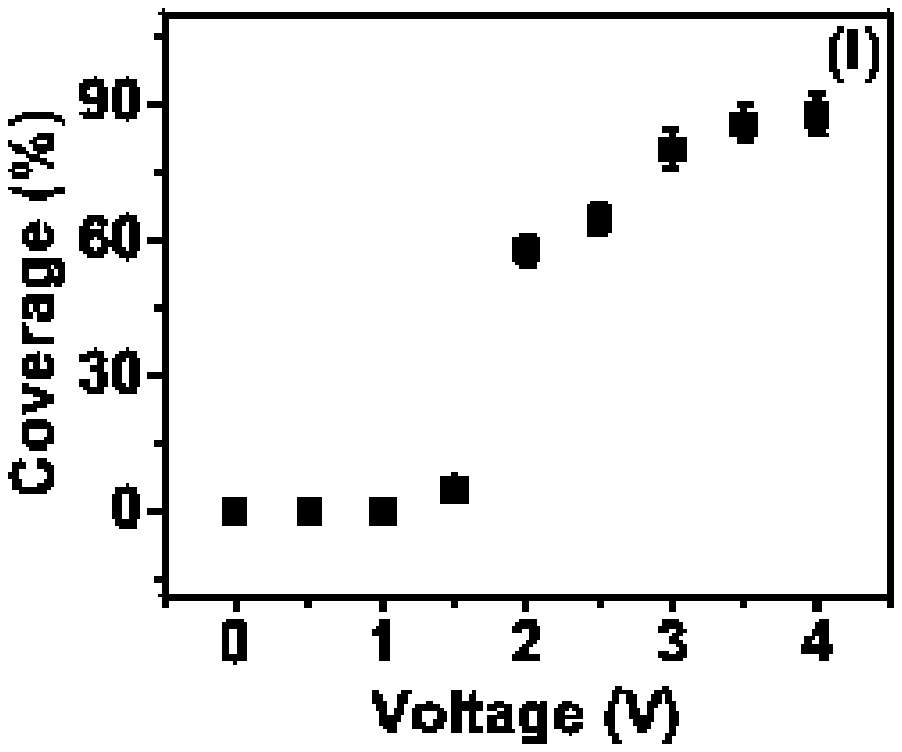}
\includegraphics[height=55mm]{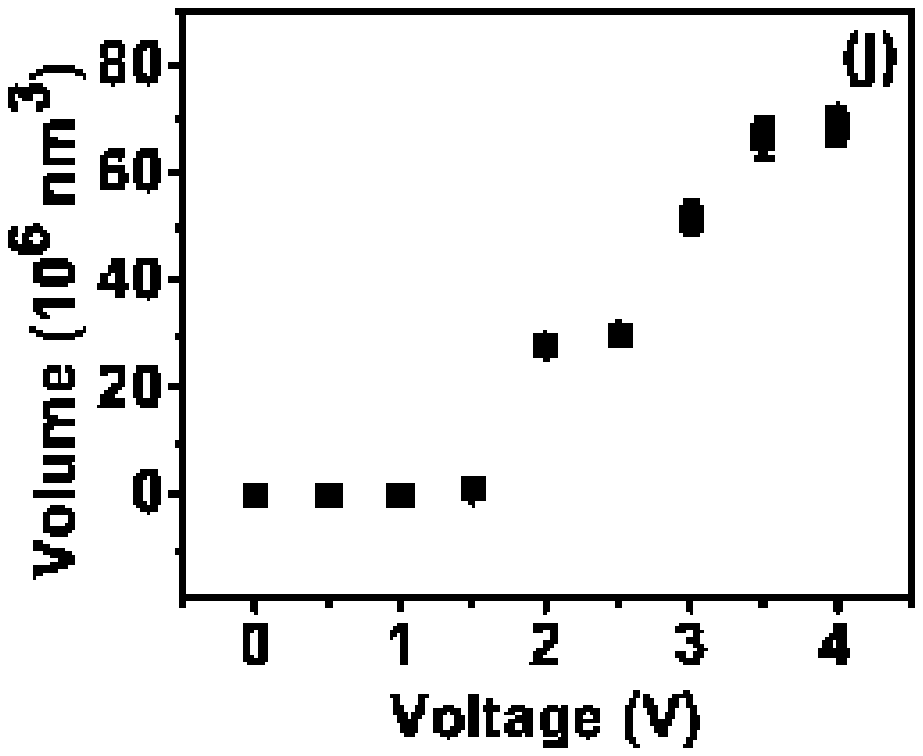}
\vspace{1mm}\\
\caption{ \label{2} (color) AFM (tapping mode) topography images
of HOPG surface (under water) as cathode at different voltages:
(a) $1\,{\rm V}$, (b) $1.5\,{\rm V}$, (c) $2\,{\rm V}$, (d)
$2.5\,{\rm V}$, (e) $0\,{\rm V}$, (f) $3\,{\rm V}$, (g) $3.5\,{\rm
V}$, and (h) $4\,{\rm V}$ (height range: (a-b) $42\,{\rm nm}$,
(c-e) $50.6\,{\rm nm}$, (f) $61.2\,{\rm nm}$, (g-h) $115.5\,{\rm
nm}$). The scanning time per image is $8.5\,{\rm min}$ and the
images are taken in a sequence from (a) to (h). Hydrogen
nanobubbles are produced on the surface. When the nanobubble
coverage is low, the atomic steps traversing the HOPG surface are
visible. The formation of nanobubbles increases tremendously when
increasing the voltage from $1.5$ to $2\,{\rm V}$. In (e) the
voltage has been switched off, while the nanobubbles remain
stable. In (f-h), at higher voltages nanobubbles cover the entire
surface with much larger individual sizes. Nanobubbles growing
(marked by arrow 1 and 3) or detaching (marked by arrow 2) are
observed. The dependence of nanobubble coverage and volume upon
the applied voltage is shown as plot (i) and (j) (error bar $\pm$
5\%), respectively. With increasing voltage, more hydrogen
molecules are produced at the cathode (HOPG surface), enhancing
the local gas concentration. This results in more and larger
nanobubbles, as revealed by (i) and (j).}
\end{figure*}

\begin{figure*}
\includegraphics[height=40mm]{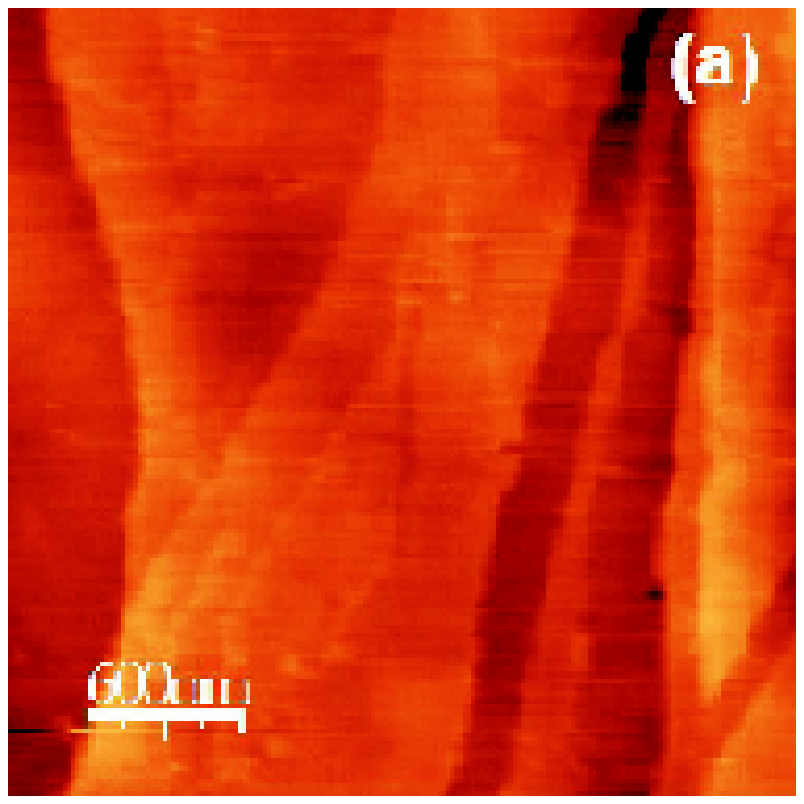}
\includegraphics[height=40mm]{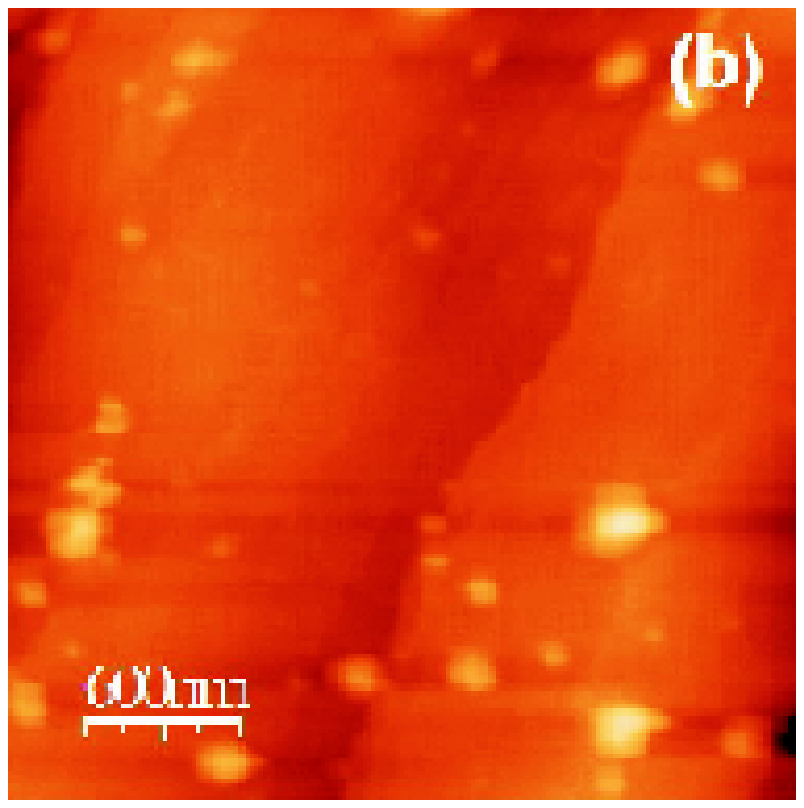}
\includegraphics[height=40mm]{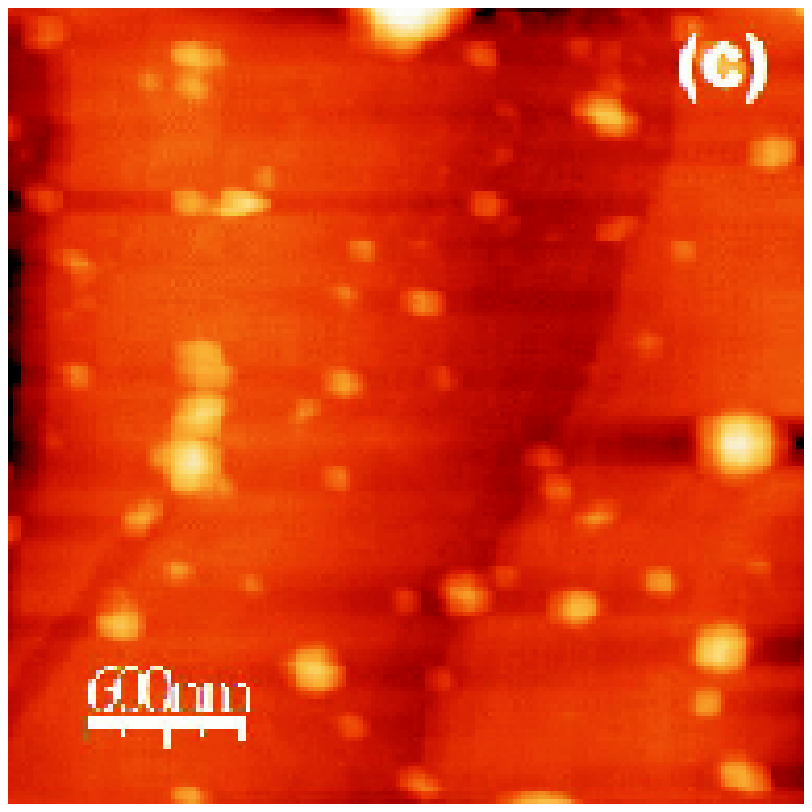}
\includegraphics[height=40mm]{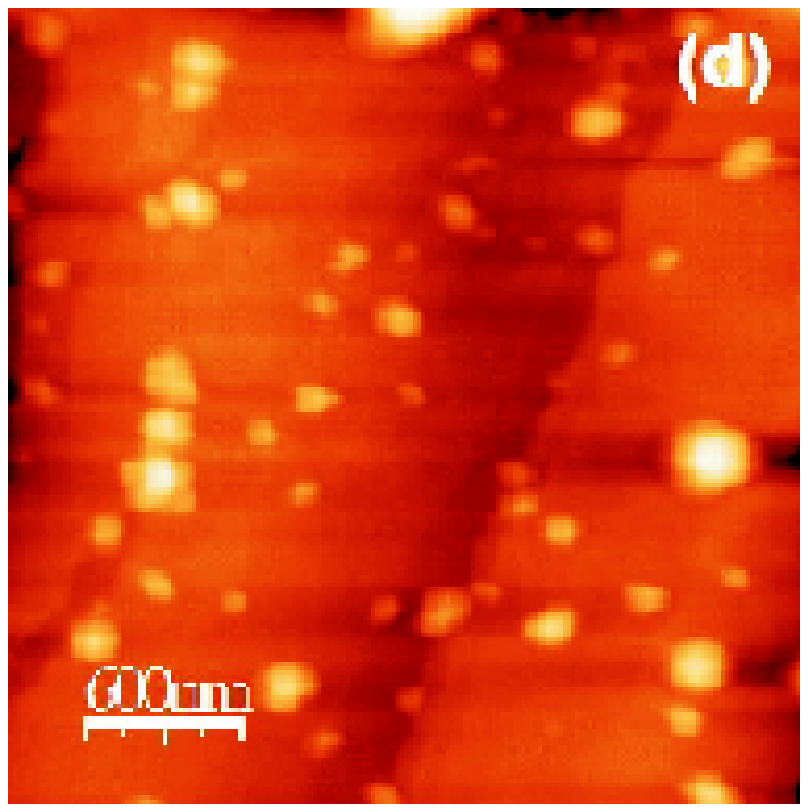}
\includegraphics[height=40mm]{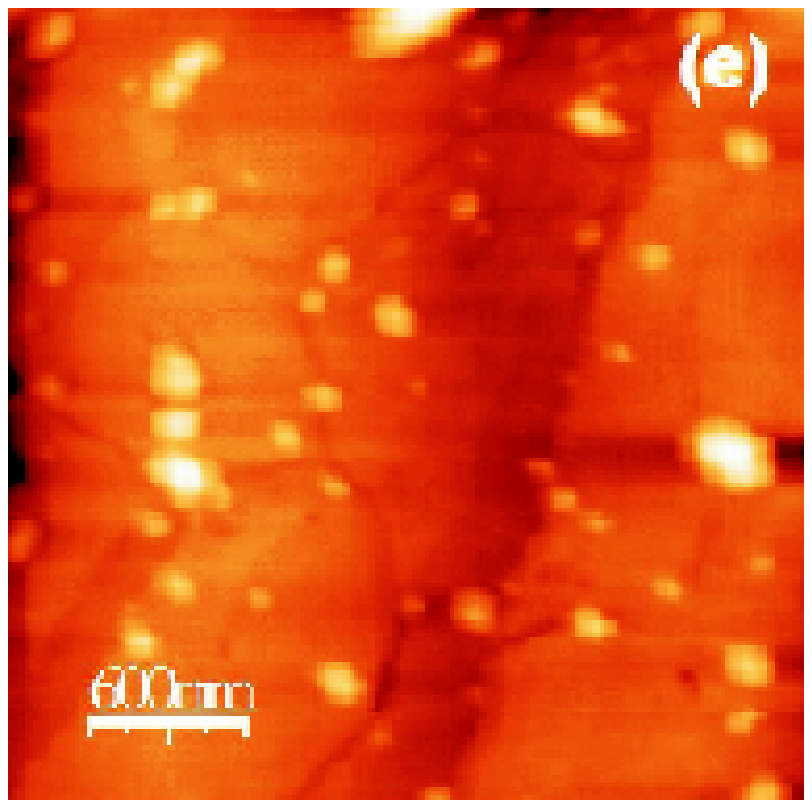}
\includegraphics[height=40mm]{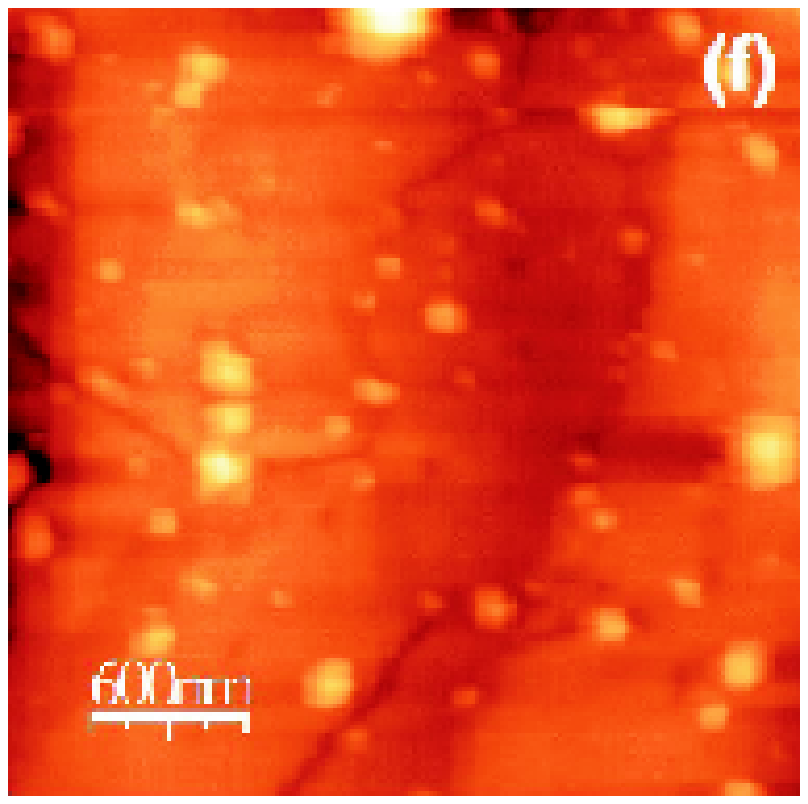}
\includegraphics[height=40mm]{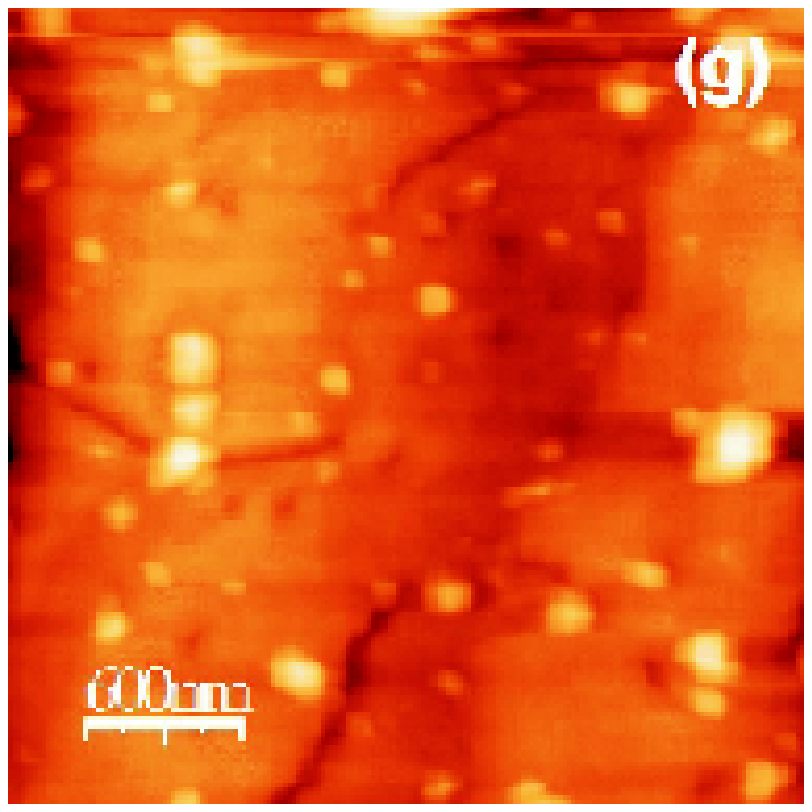}
\includegraphics[height=55mm]{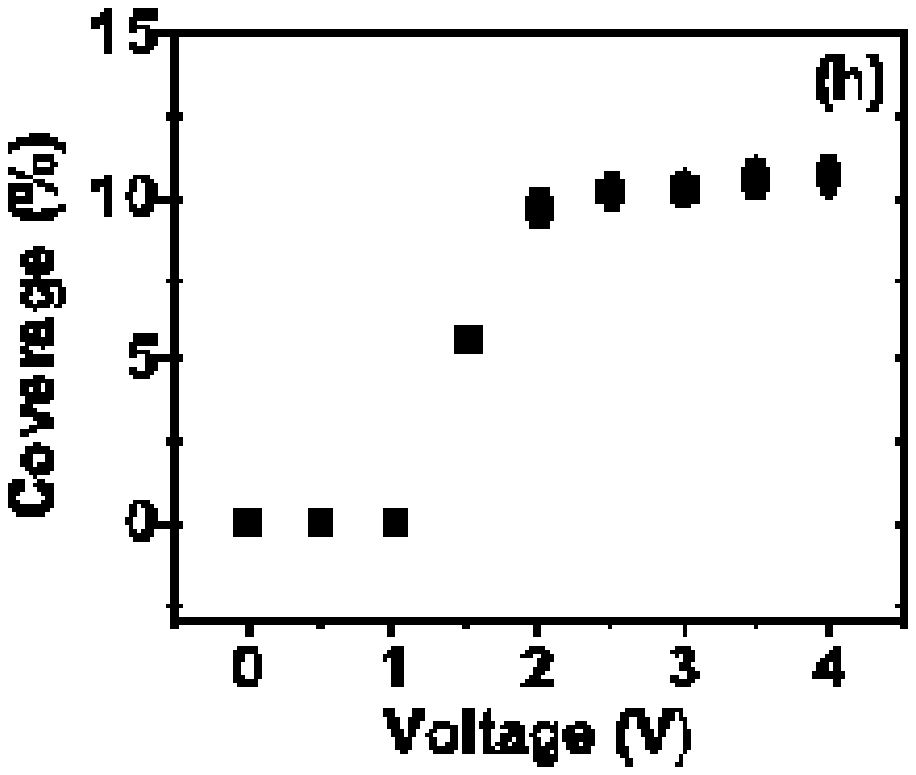}
\includegraphics[height=55mm]{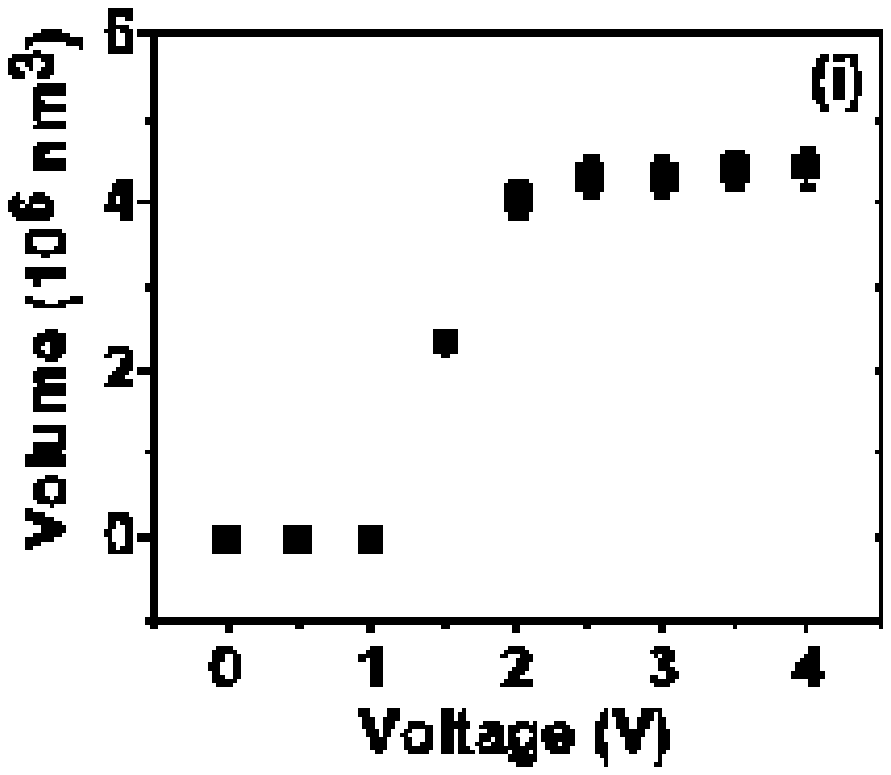}
\vspace{1mm}\\
\caption{ \label{3} (color) AFM (tapping mode) topography images
of HOPG surface (under water) as anode at different voltages: (a)
$1\,{\rm V}$, (b) $1.5\,{\rm V}$, (c) $2\,{\rm V}$, (d) $2.5\,{\rm
V}$, (e) $3\,{\rm V}$, (f) $3.5\,{\rm V}$, and (g) $4\,{\rm V}$
(height range: (a) $12\,{\rm nm}$, (b-g) $35\,{\rm nm}$). Again,
the atomic steps of HOPG surface are visible. Images are recorded
continuously from (a) to (g), with a scanning time of $8.5\,{\rm
min}$ per image. Nanobubbles (oxygen) are formed on the surface.
Comparing to the hydrogen case in Figure 2, the number and volume
of the produced oxygen nanobubbles is much smaller. This is
presumably due to: i) the considerable difference of solubility in
water between oxygen and hydrogen (oxygen's solubility is $\sim2$
times higher than hydrogen's at 20 $^\circ$C); ii) the difference
in the production rate during the electrolysis, ${\rm H}_2:{\rm
O}_2=2:1$. Plot (h) and (i) show coverage and volume of the
nanobubbles as a function of the imposed voltage, respectively
(error bar $\pm$ 5\%).}
\end{figure*}

\begin{figure*}
\includegraphics[height=40mm]{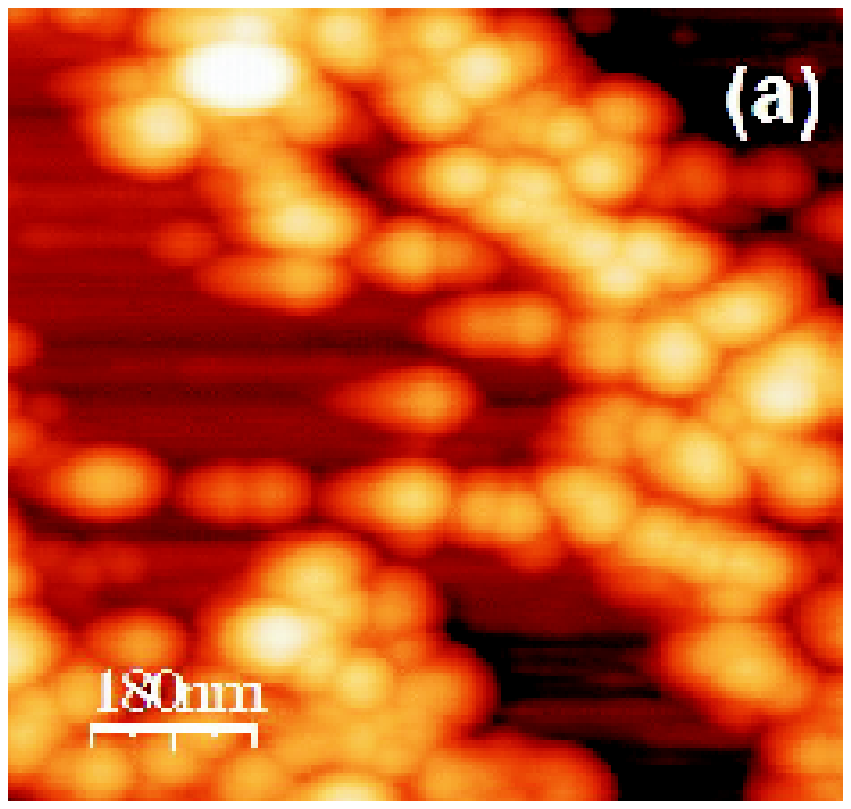}
\includegraphics[height=40mm]{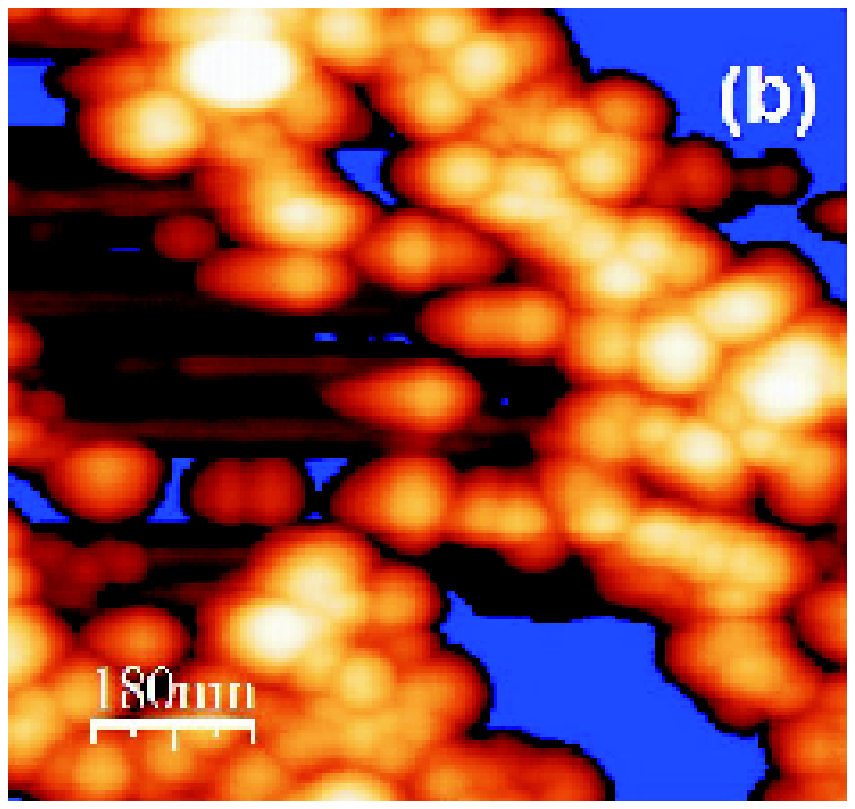}
\includegraphics[height=40mm]{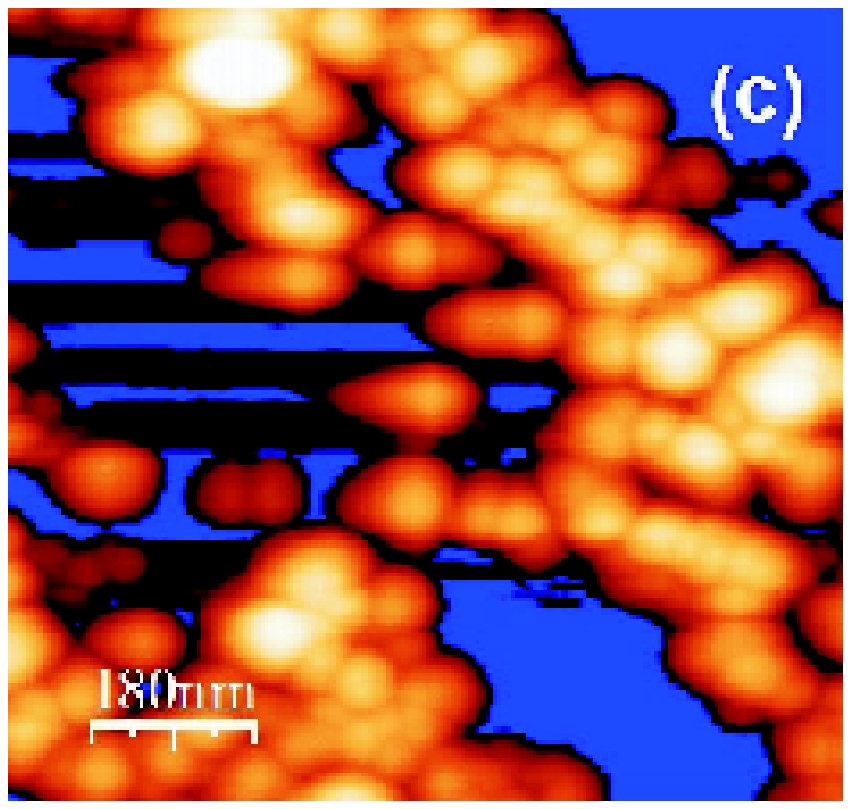}
\includegraphics[height=40mm]{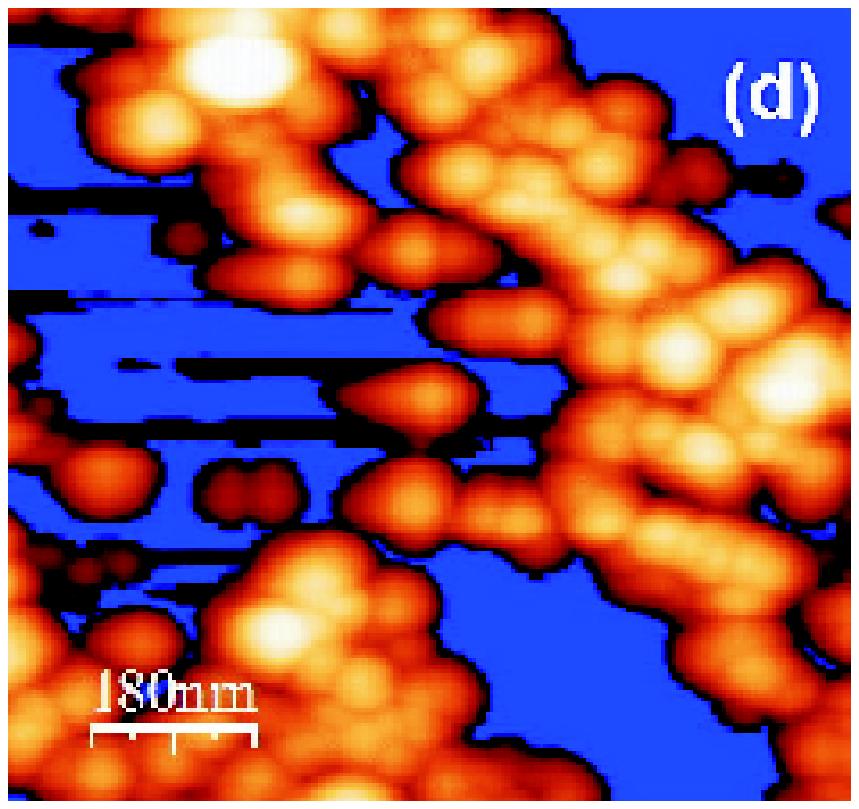}
\includegraphics[height=40mm]{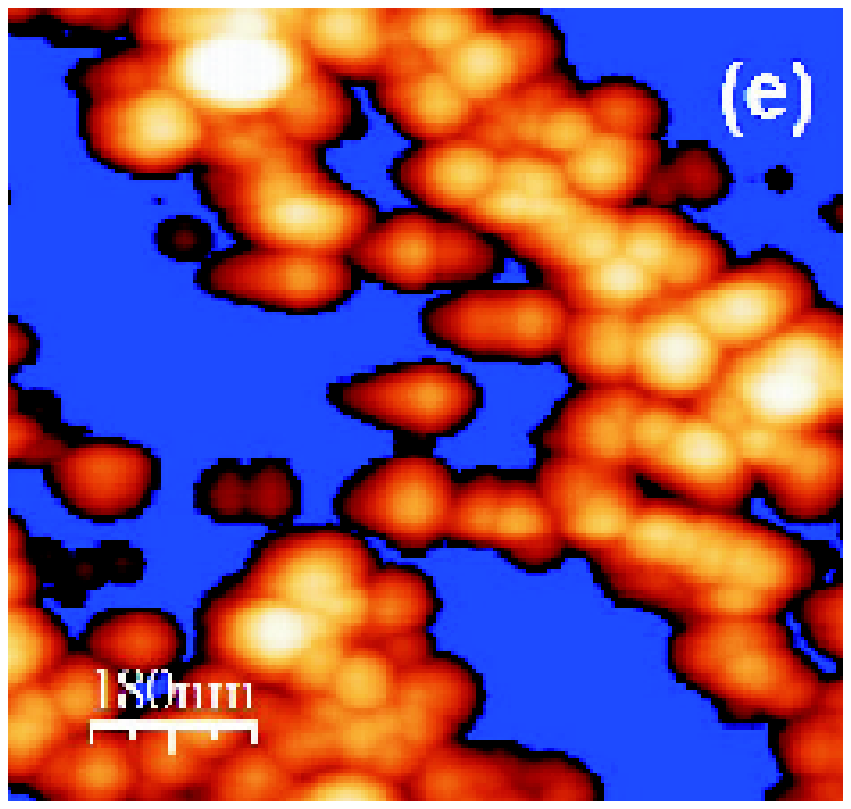}
\includegraphics[height=40mm]{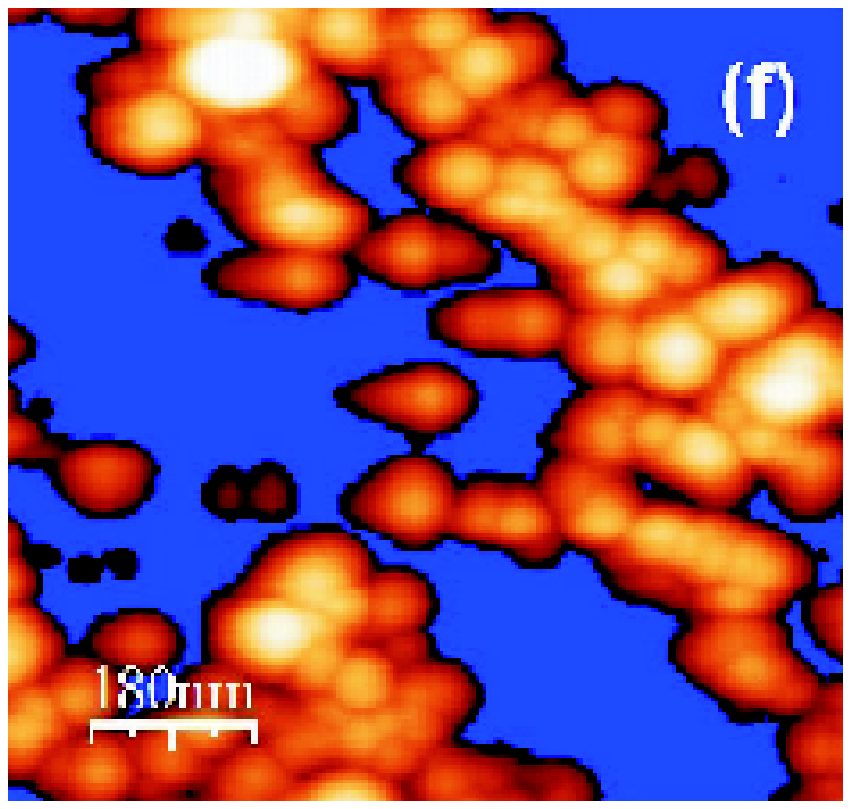}
\includegraphics[height=40mm]{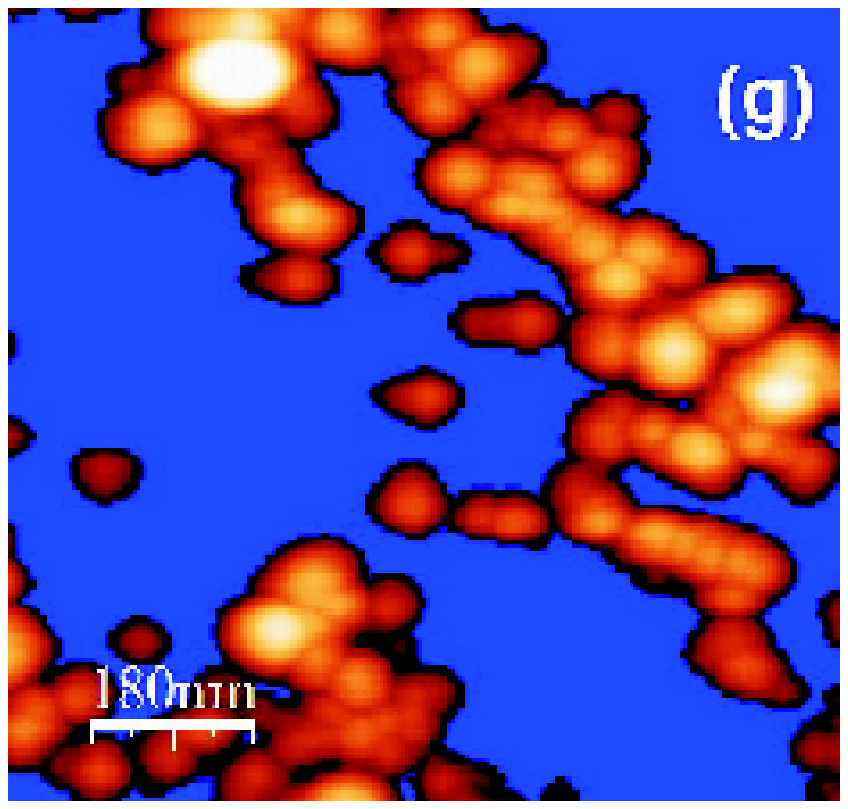}
\includegraphics[height=35mm]{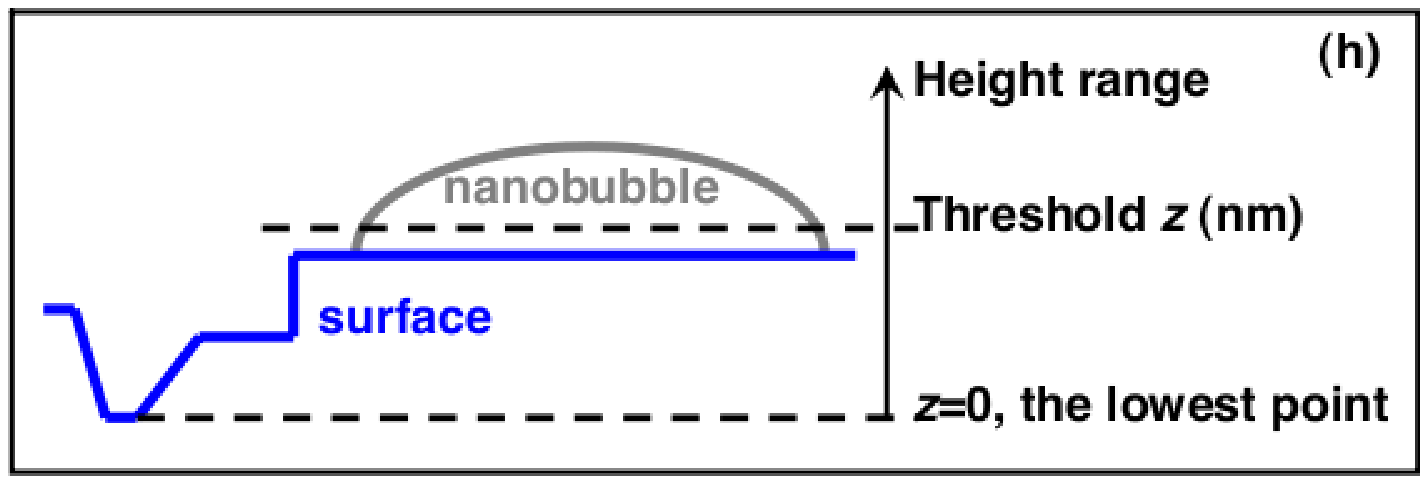}
\includegraphics[height=50mm]{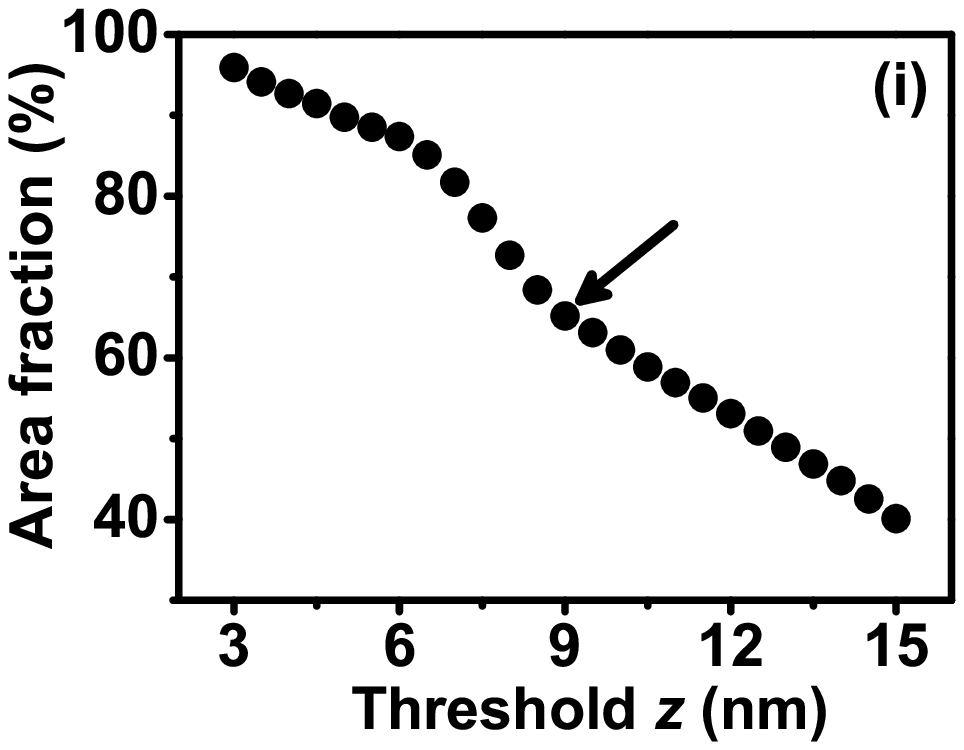}
\vspace{1mm}\\
\caption{ \label{4}(color) Tapping mode AFM topography images
(height range $27.2\,{\rm nm}$) of hydrogen surface nanobubbles,
when different thresholds $z$ are applied for the identification
of surface nanobubbles: (a) $z$=0 nm, (b) $z$=6 nm, (c) $z$=7 nm,
(d) $z$=8 nm, (e) $z$=9 nm, (f) $z$=10 nm, and (g) $z$=14 nm.
Sketch (h) describes the principle. Areas below this threshold are
mashed as blue, while areas above, depending on the height, as
yellowish. The fraction of the latter area is shown in (i) as
function of the threshold $z$. That curve shows a pronounced
shape. We take the end of the straight shape region (see arrow and
$z$=9 nm) towards smaller $z$ as estimate for the nanobubble
coverage.}
\end{figure*}

\begin{figure*}
\includegraphics[height=80mm]{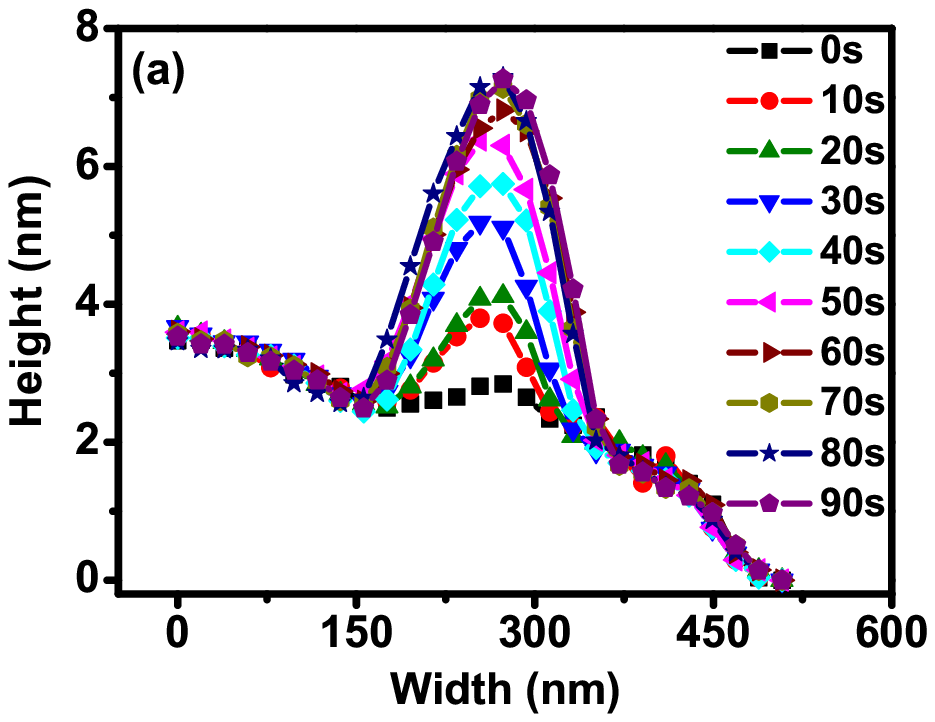}
\includegraphics[height=85mm]{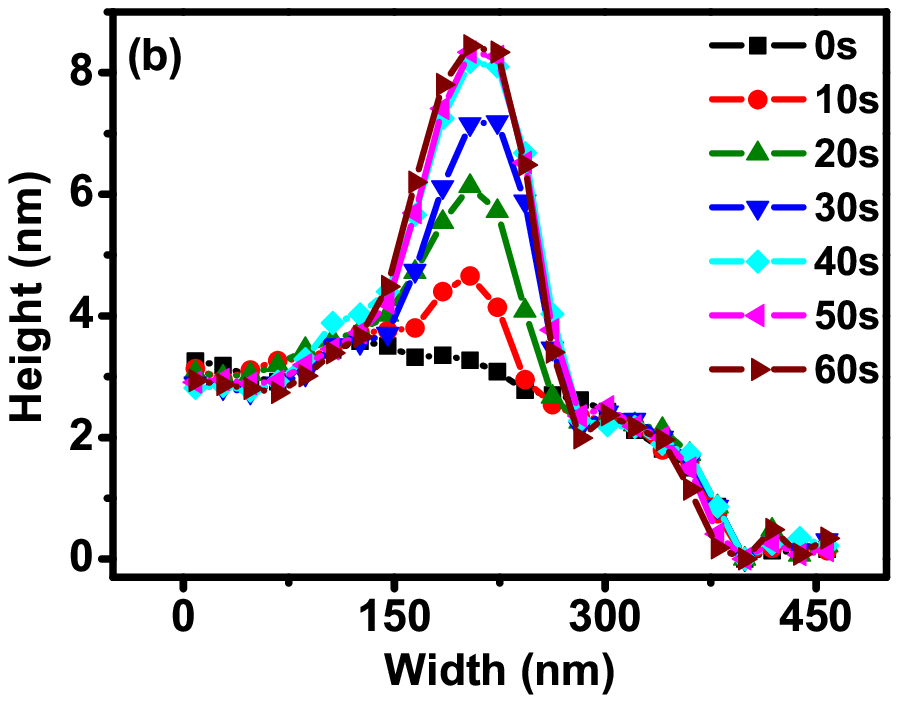}
\vspace{0mm}\\
\caption{ \label{5} (color) (a) shows the real-time profiles of a
nanobubble on HOPG surface (as cathode) at 1 V, with time interval
of 10 sec. Another example at 2 V is shown in (b). By means of
electrolysis of water, nanobubbles form on the surface and
subsequently grow. In (a) the growth terminates after 70 sec while
in (b) this occurs already after 40 sec. The nanobubbles then
remain stable. The plots also reveal that the nanobubbles grow
with a higher rate in height rather than in width.}
\end{figure*}

\begin{figure*}
\includegraphics[height=50mm]{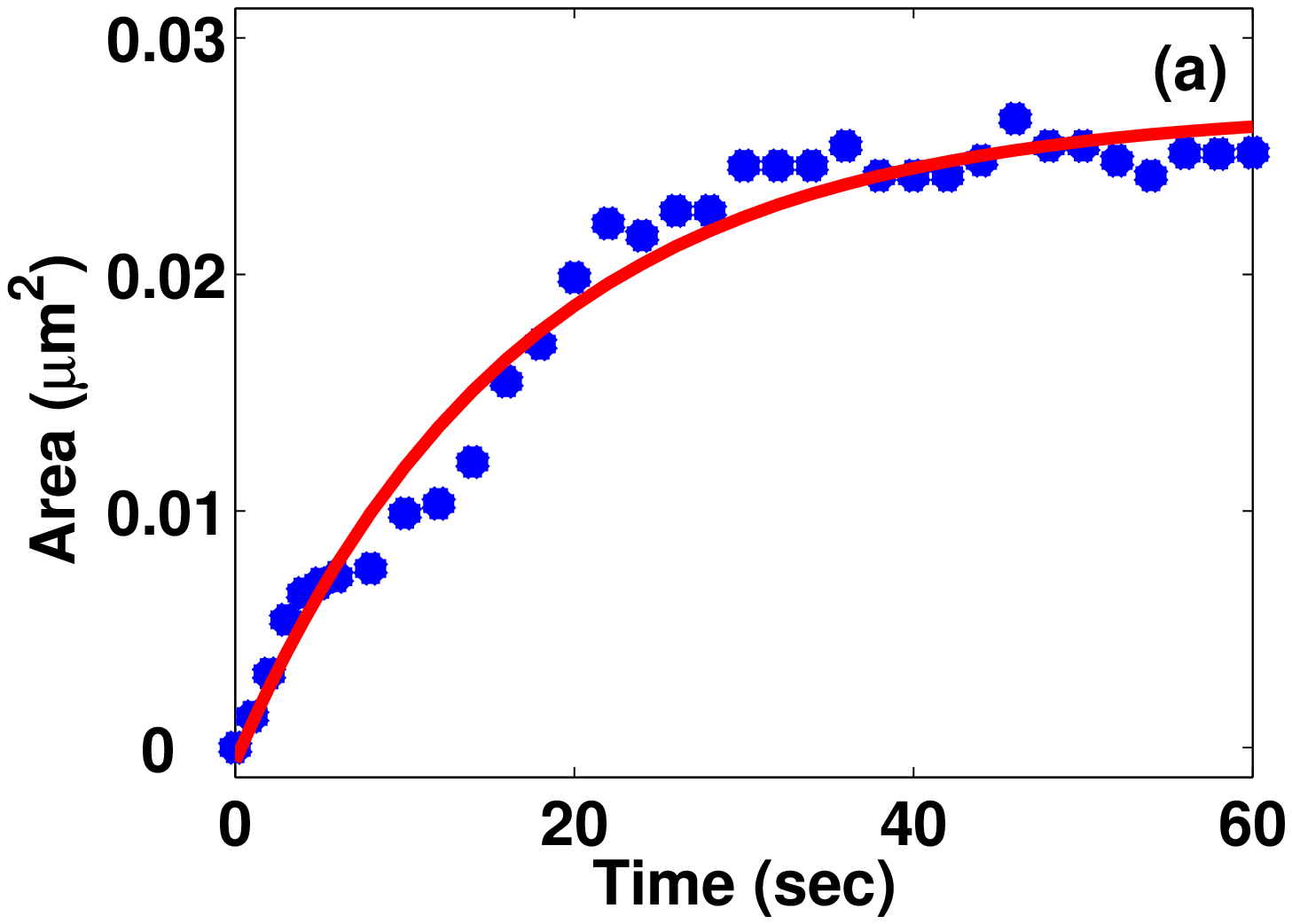}
\includegraphics[height=50mm]{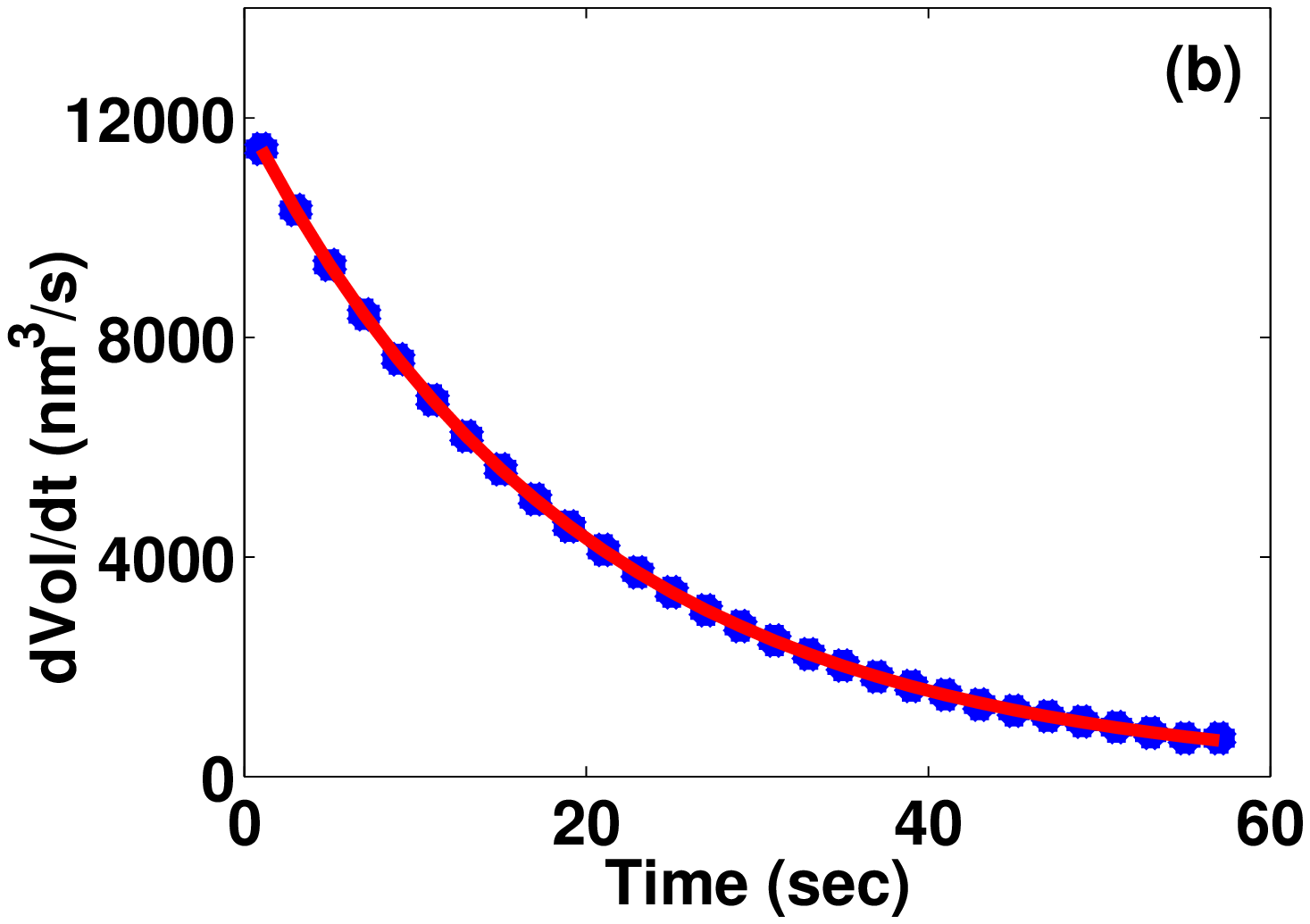}
\includegraphics[height=45mm]{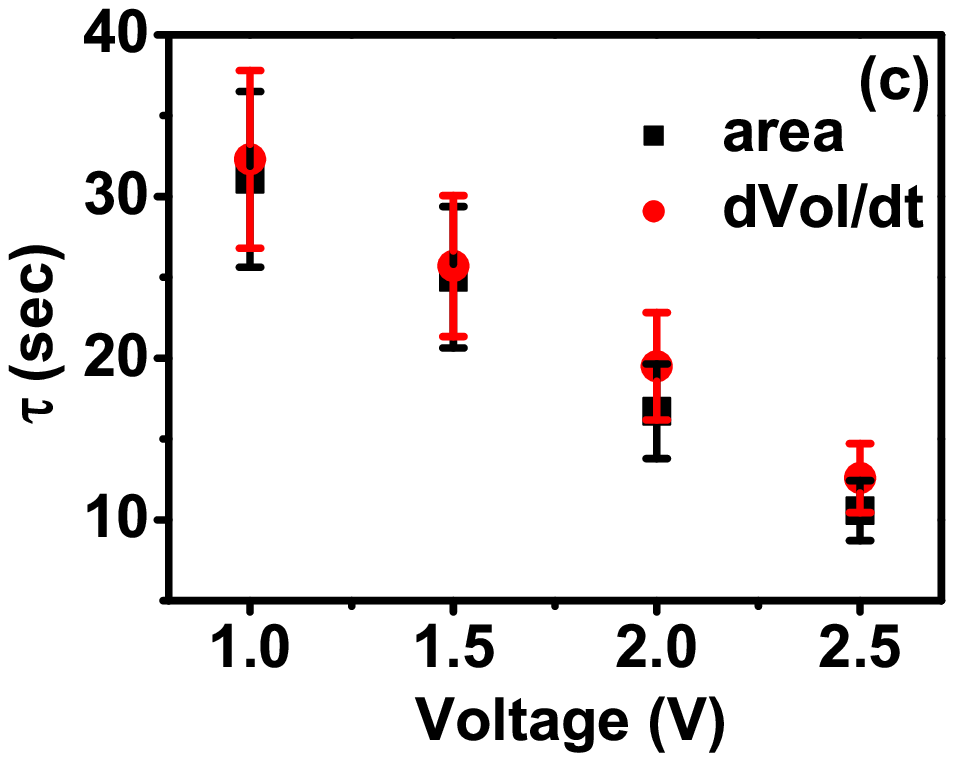}
\vspace{0mm}\\
\caption{ \label{6} (color) nanobubble area (a) and volume growth
rate (b) plots (blue dots). The red curves are fits of an
exponential function $X=X_\infty+(X_0-X_\infty)e^{-t/\tau}$, where
$X$ is either the area or the volume growth rate. These fits allow
to define a characteristic timescale $\tau$. Values of the time
constant $\tau$ extracted from the fits are exhibited as a
function of the applied voltage, as shown in (c). The timescales
of the area evolution (black square) and the volume growth rate
evolution (red dot) show a good correlation at all voltages. This
observation suggests two possible ways how the electrolytic gas is
produced on the surface: i) the gas emerges at the whole surface
of the nanobubbles and correspondingly the whole surface of the
nanobubbles should be charged by electrons; ii) the gas emerges at
the electrode surface and then diffuses through the nanobubble
surface.}
\end{figure*}

\begin{figure*}
\includegraphics[height=50mm]{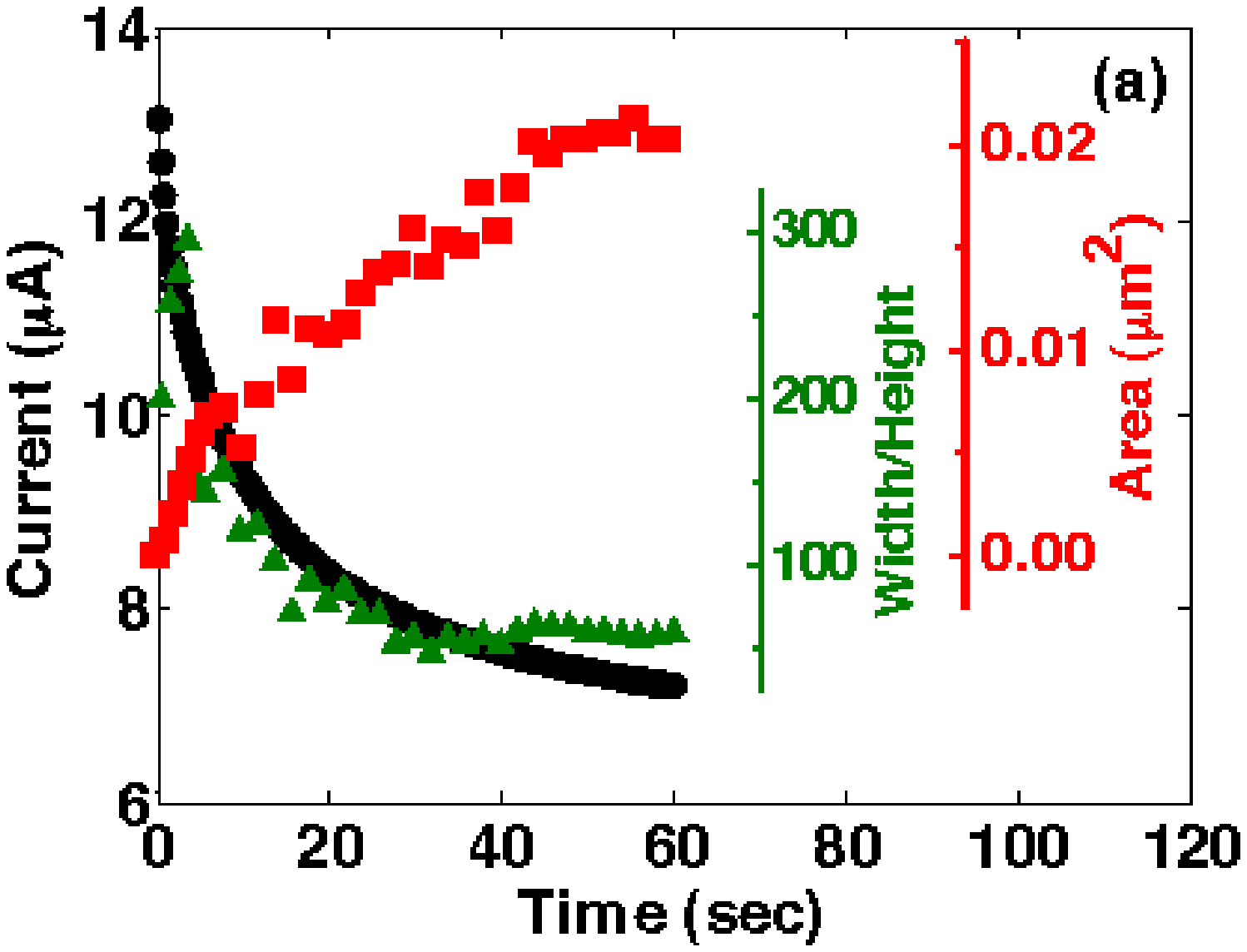}
\hspace{2mm}
\includegraphics[height=50mm]{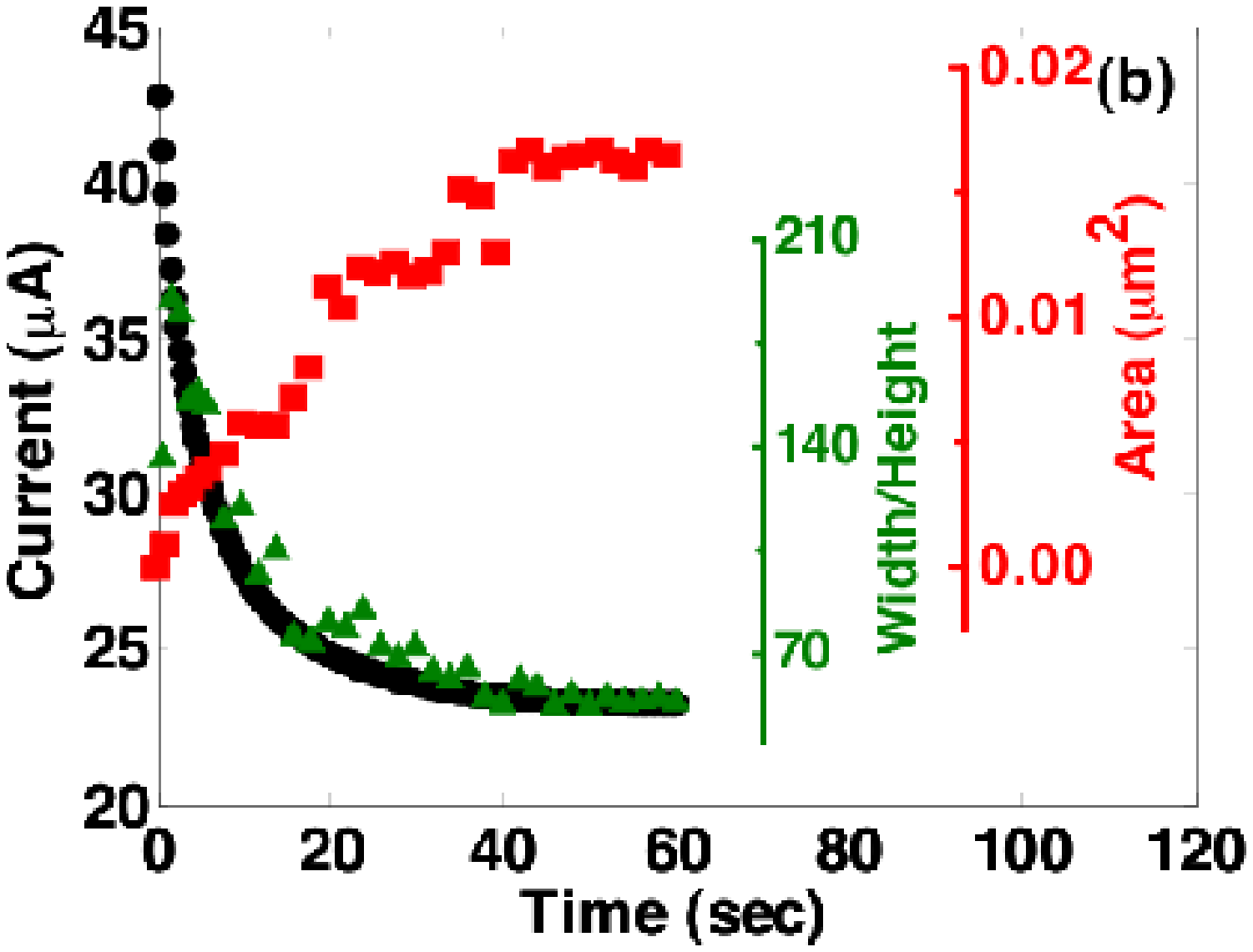}
\includegraphics[height=50mm]{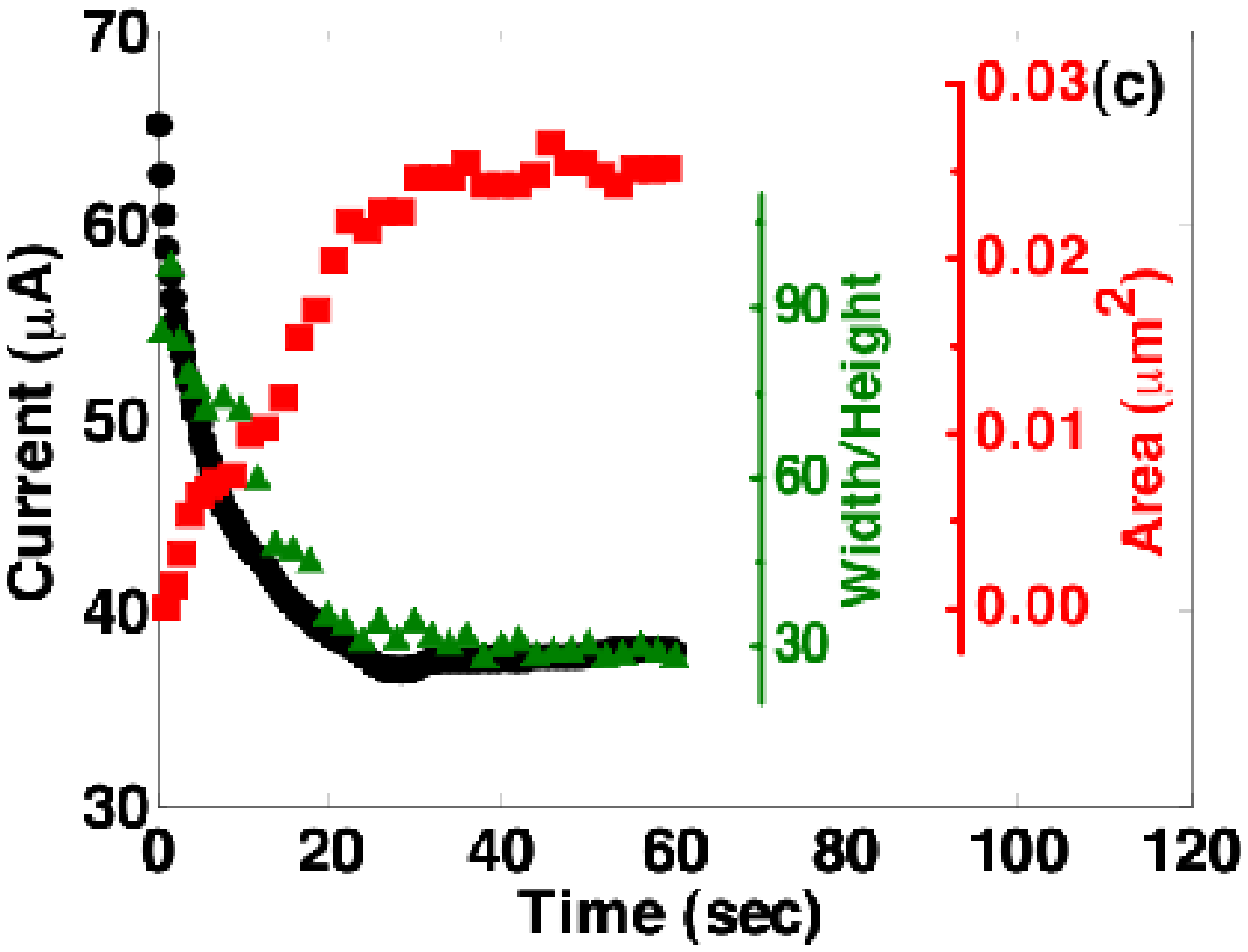}
\hspace{2mm}
\includegraphics[height=50mm]{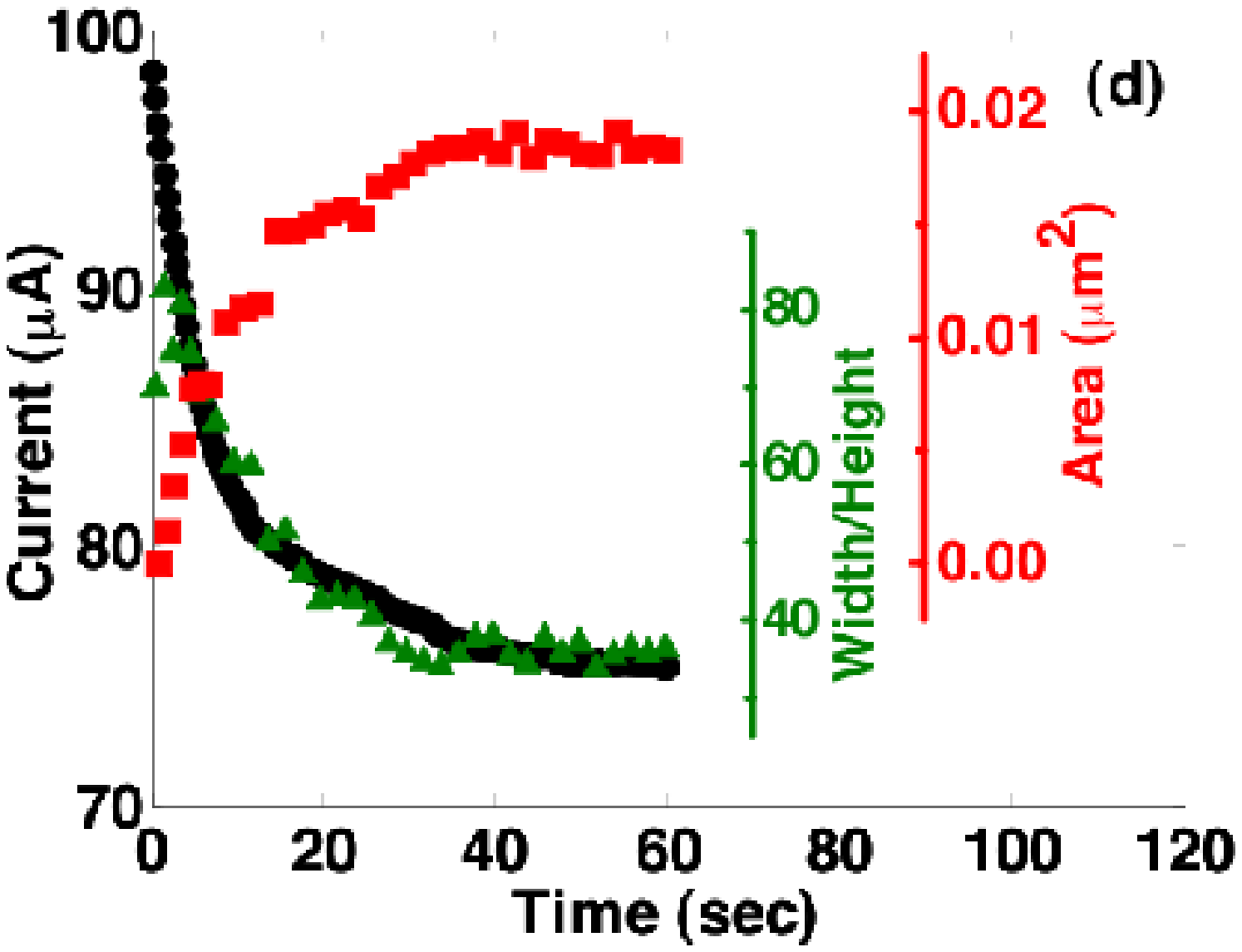}
\vspace{2mm}\\
\caption{ \label{7} (color) Graphs showing current, nanobubble
area, and nanobubble aspect ratio (width/height) as a function of
time at (a) 1 V, (b) $1.5$ V, (c) 2 V, and (d) $2.5$ V on HOPG as
cathode. At each voltage, the three plots are recorded
simultaneously. The nanobubble development and the current decay
show a clear correlation. Interestingly, the current and the
nanobubble aspect ratio (green triangle) decrease in the same
manner. The aspect ratio plot indicates that nanobubbles initially
form in an ultrathin-film form and then accumulate with a higher
rate in vertical direction rather than in horizontal; this is
consistent with the finings shown in Figure 5.}
\end{figure*}

\begin{figure*}
\includegraphics[height=60mm]{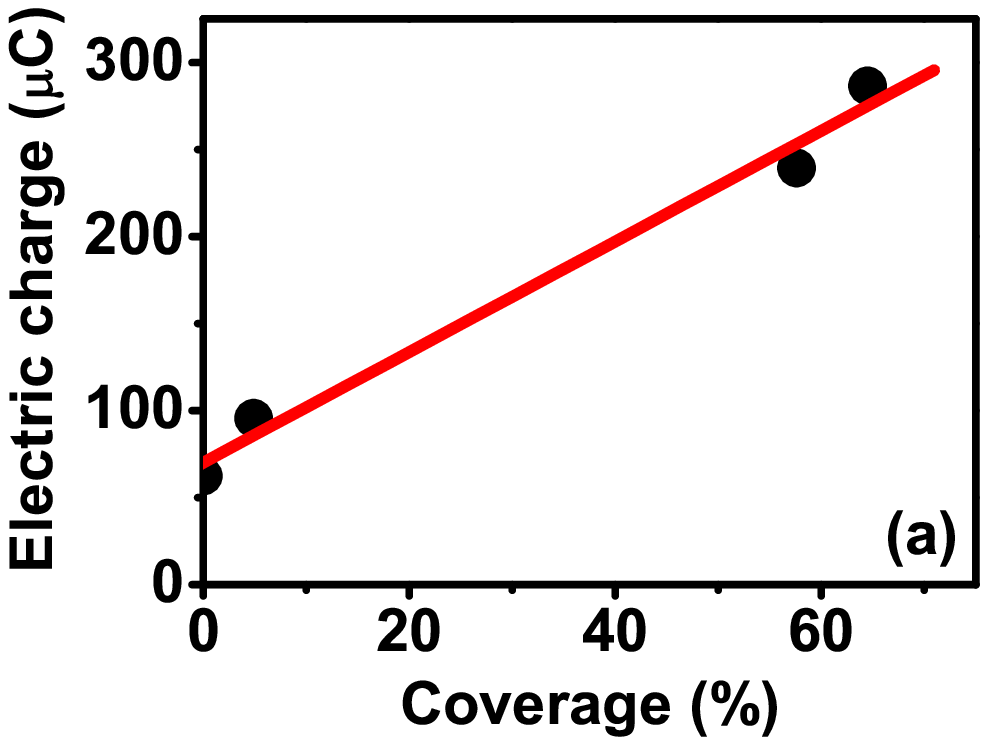}
\includegraphics[height=57mm]{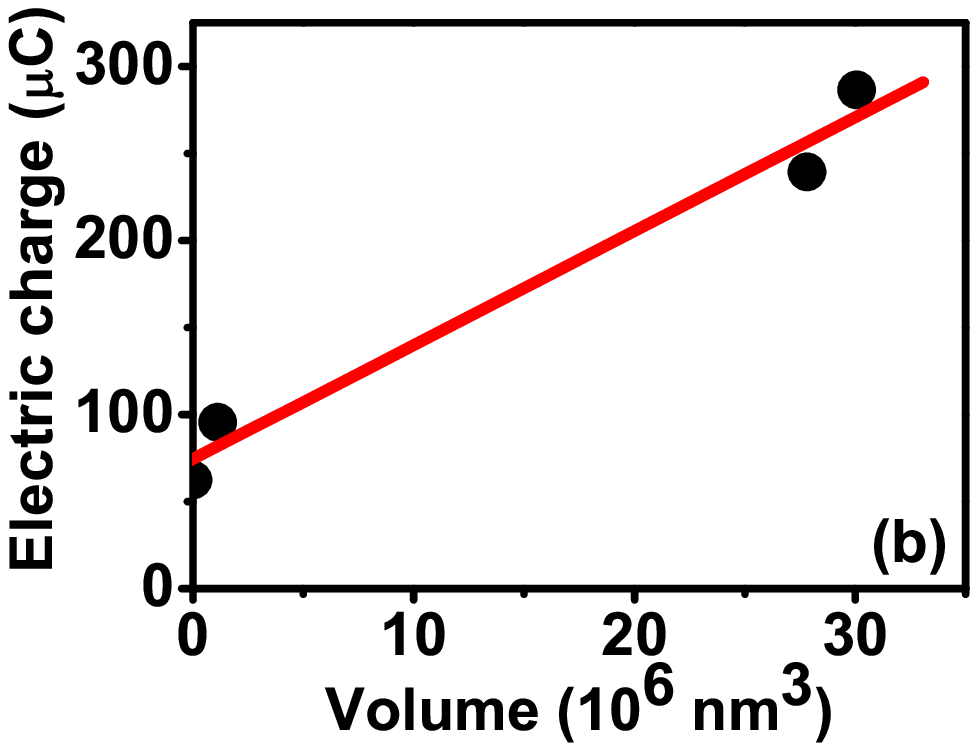}
\vspace{0mm}\\
\caption{ \label{8} (color) The amount of the excess electric
charge in equilibrium (after 60 sec) is estimated for each
voltage, namely 1, $1.5$, 2, and $2.5$ V. It is plotted versus the
nanobubble coverage (a) and the volume (b). The red lines are
linear fits. Note the offset of the linear fits: A finite amount
of charge is needed before nanobubbles are produced, presumably in
order to build up dielectric layers at the interface.}
\end{figure*}

\begin{figure*}
\includegraphics[height=50mm]{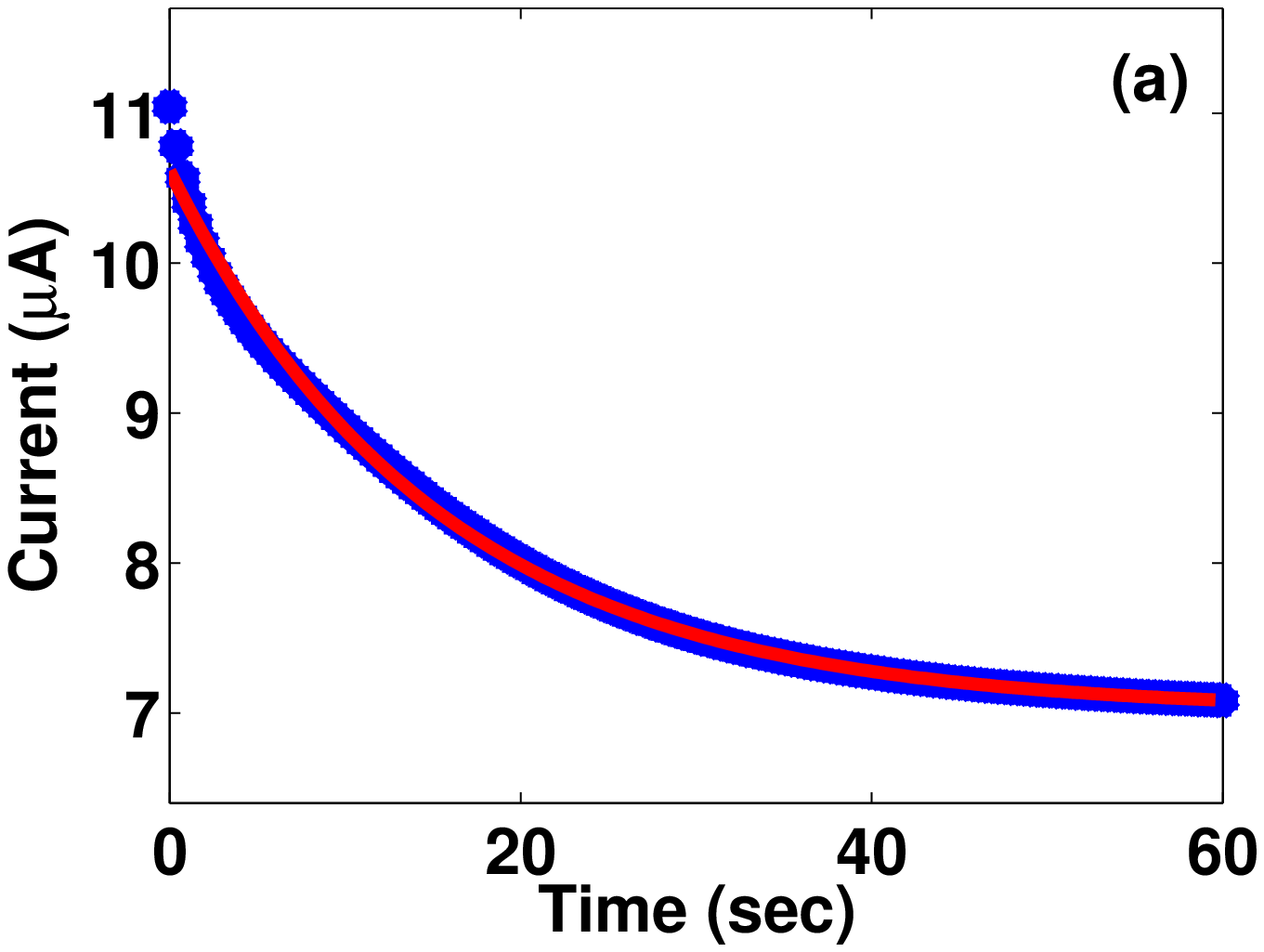}
\includegraphics[height=50mm]{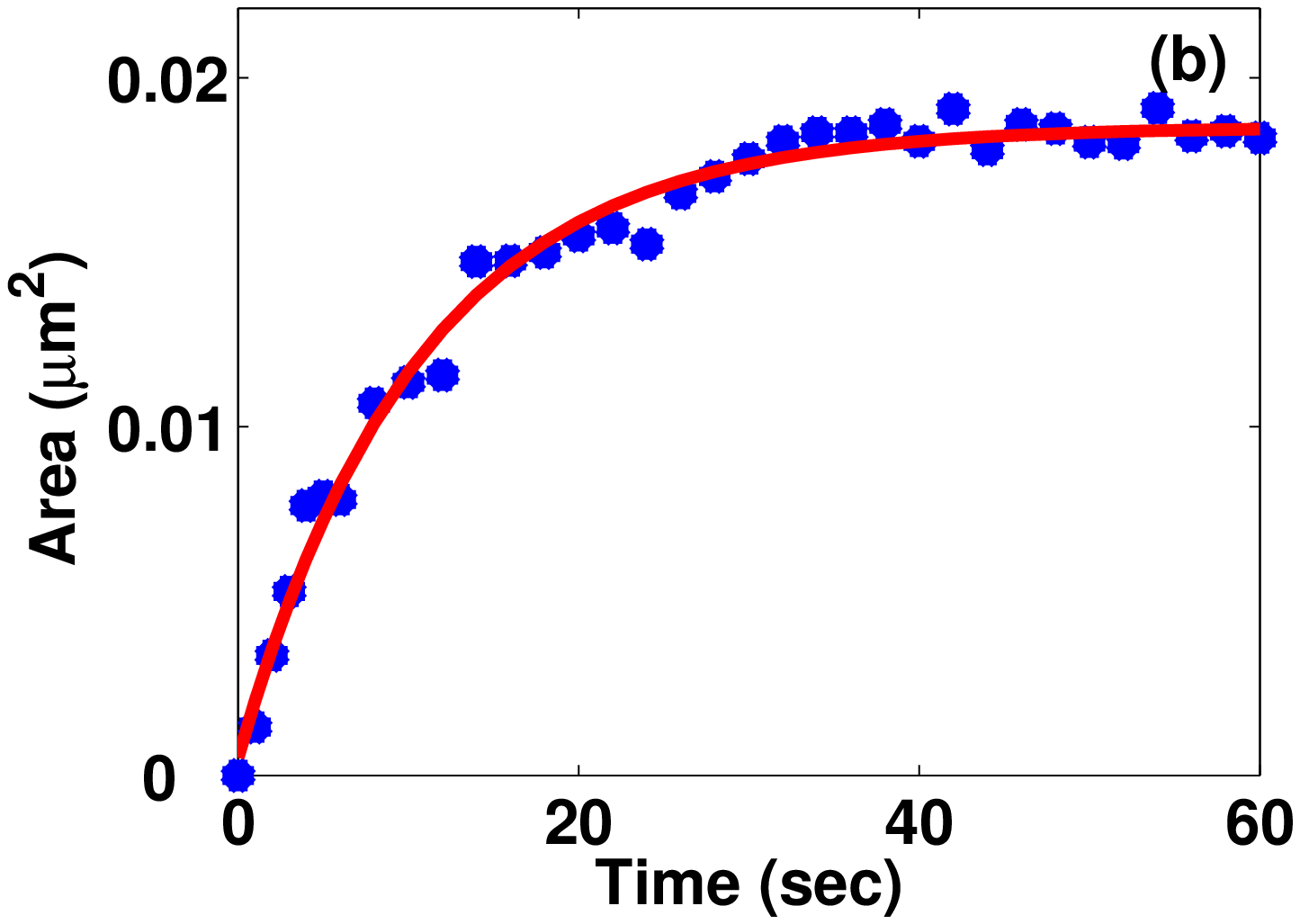}
\includegraphics[height=50mm]{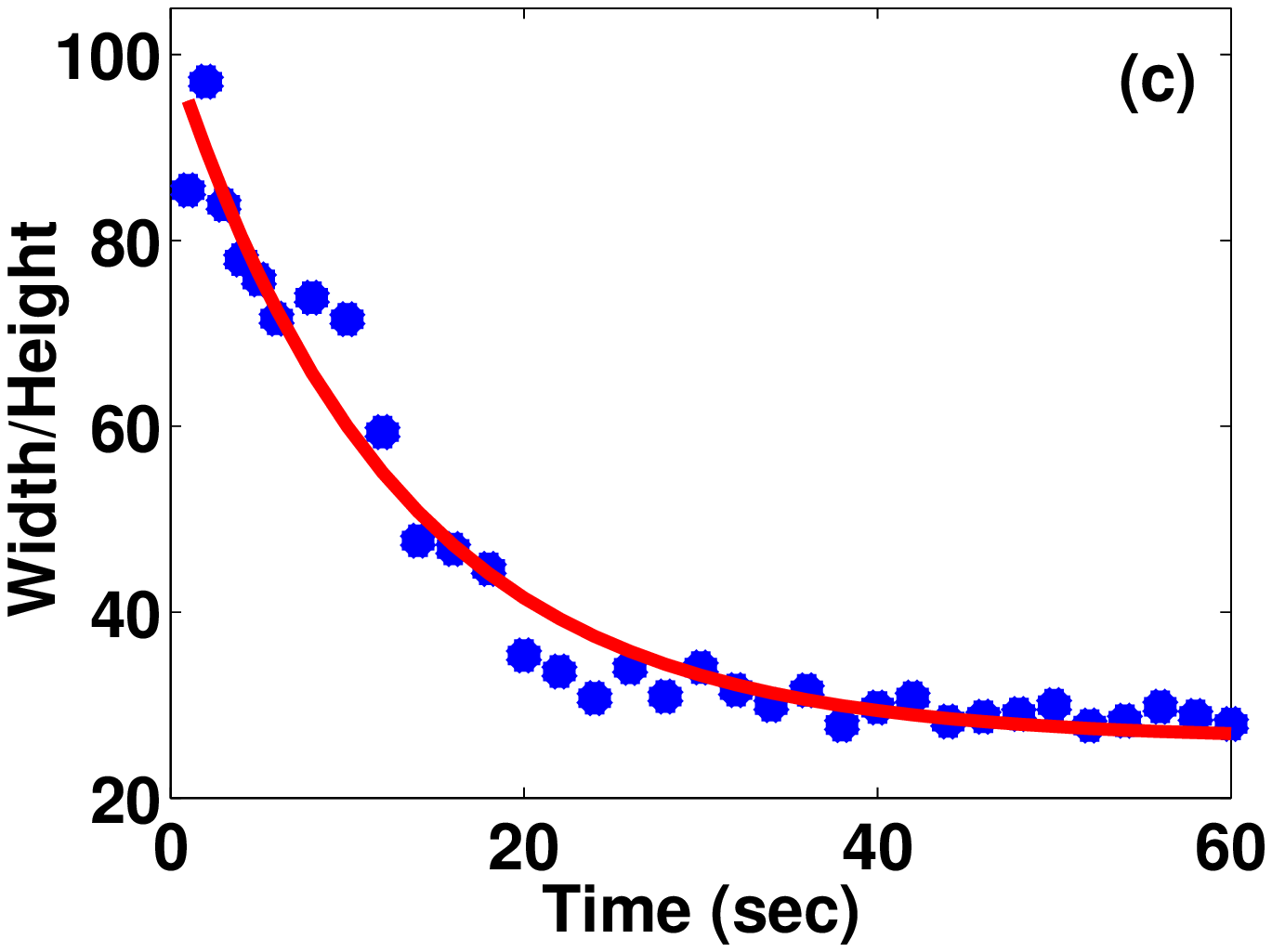}
\includegraphics[height=50mm]{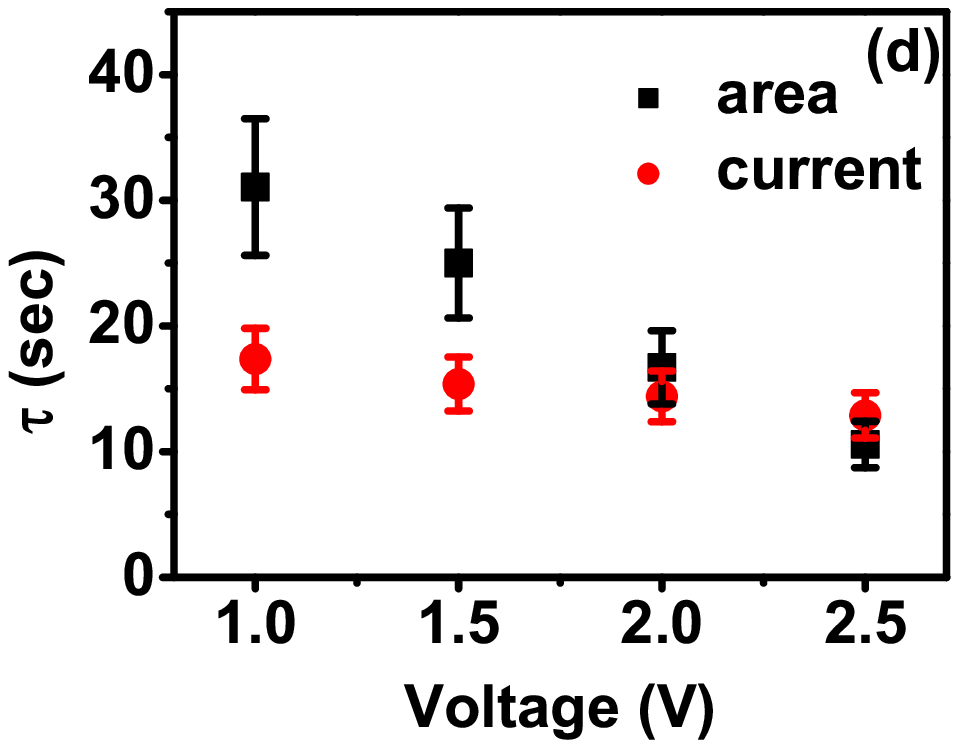}
\includegraphics[height=53mm]{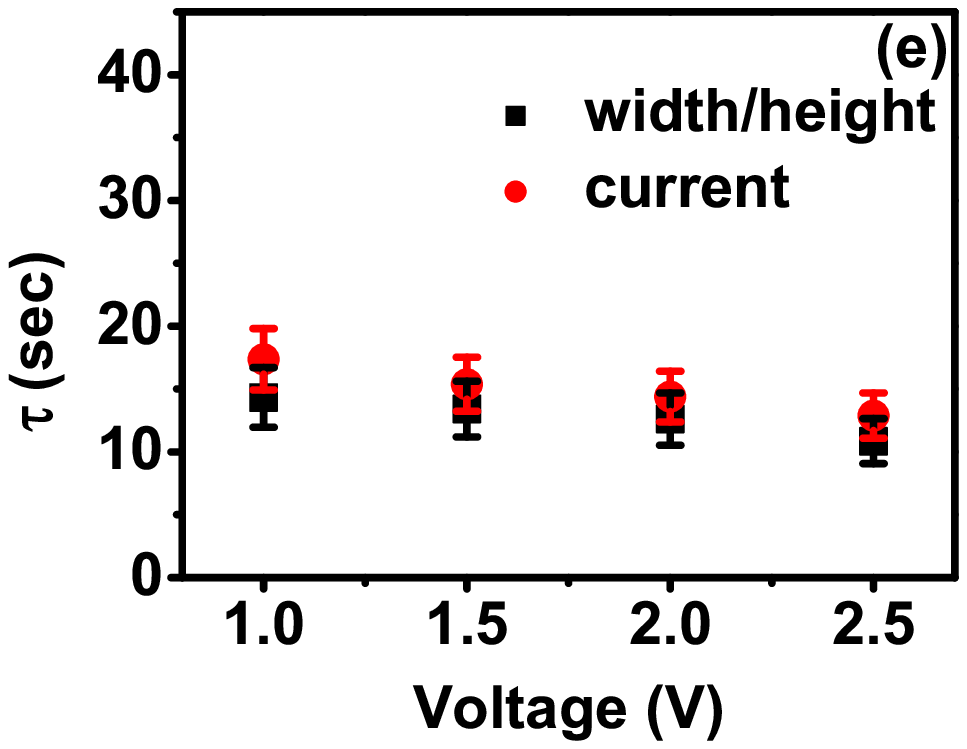}
\vspace{0mm}\\
\caption{ \label{9} (color) Exponential fits (red curves) of the
current, nanobubble area and width/height plots (blue dots) are
shown in (a), (b) and (c), respectively. Values of the time
constant $\tau$ of the fits are extracted. $\tau$ is plotted
versus voltage for area (error bar $\pm$ 17\%) and current (error
bar $\pm$ 13\%) as shown in (d), as well as for width/height
(error bar $\pm$ 16\%) and current as shown in (e). $\tau$
decreases with increasing voltage, this indicates that the
development of nanobubbles and the decay of the current take place
more rapidly at higher voltage. The $\tau$ values of the area and
the current well agree at $2$ and $2.5$ V when the nanobubble
coverage is high. At 1 and $1.5$ V, when nanobubble coverage is
rather low, the time constants of area and current deviate. Note
that the current is a global measure, whereas the area of
individual nanobubbles is a local quantity. Interestingly, the
nanobubble aspect ratio and the current always show good agreement
(e), for which we do not have proper explanations.}
\end{figure*}

\begin{figure*}
\includegraphics[height=53mm]{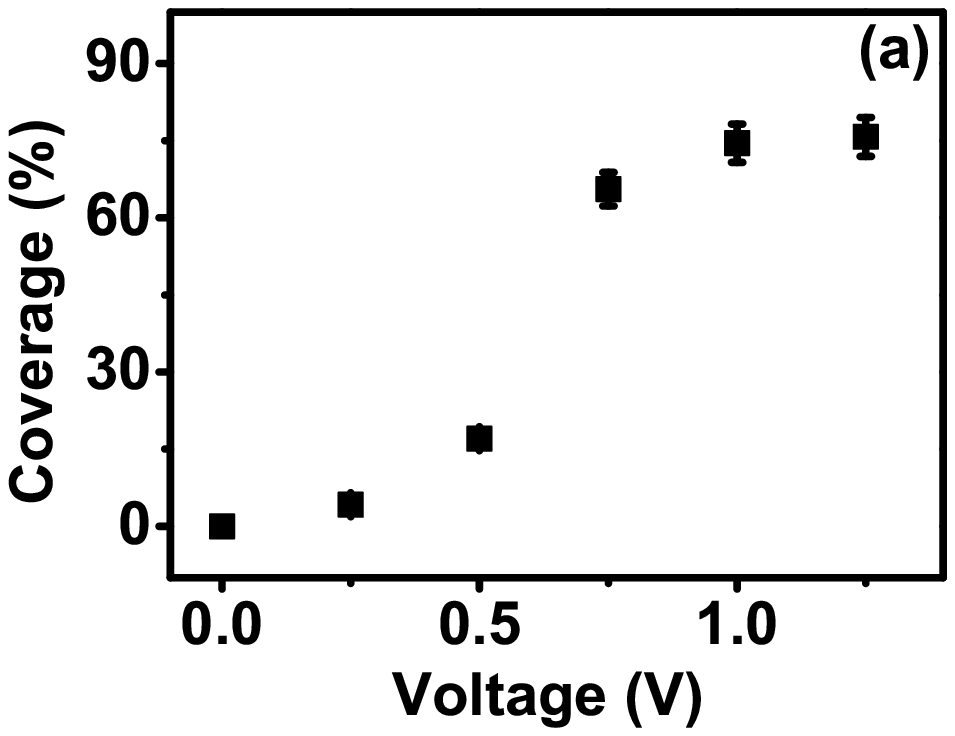}
\includegraphics[height=58mm]{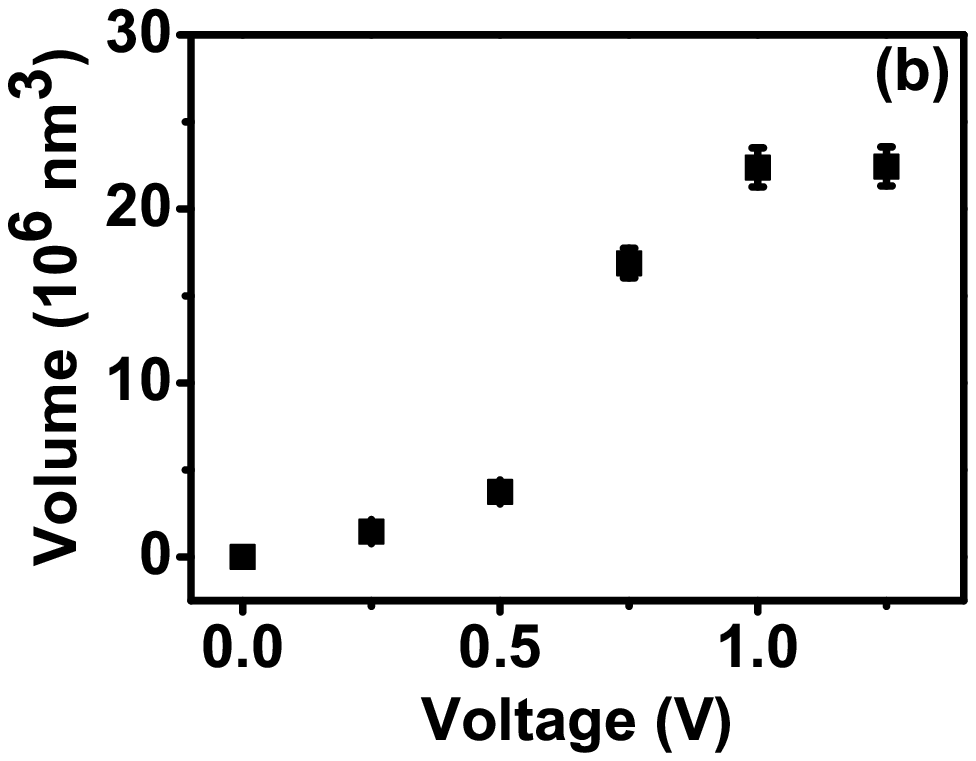}
\vspace{0mm}\\
\caption{ \label{10} Sodium chloride (NaCl) solution ($0.01$ M) is
used instead of pure water as electrolyte. Similar behavior as for
pure water is observed: with increasing voltage the formation of
hydrogen nanobubbles is enhanced (on HOPG as cathode). Coverage
and volume of the nanobubbles are related to the applied voltage,
as depicted in (a) and (b) (error bar $\pm$ 5\%) respectively. The
required effective voltage for nanobubble creation is strongly
reduced, as compared to the pure water case, as salty water has a
lower resistance.}
\end{figure*}

\begin{figure*}
\includegraphics[height=62mm]{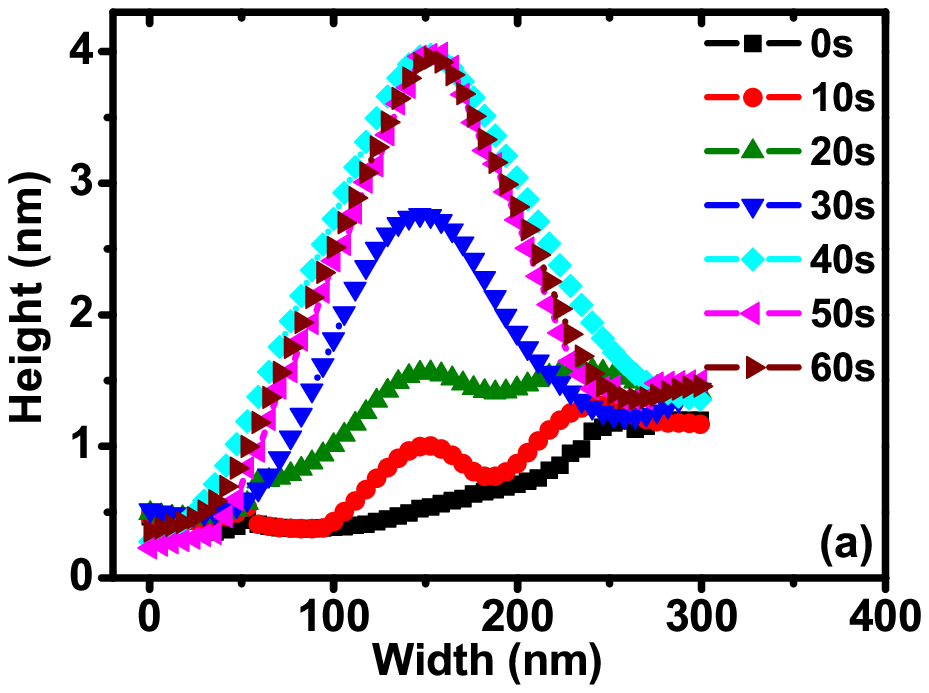}
\includegraphics[height=50mm]{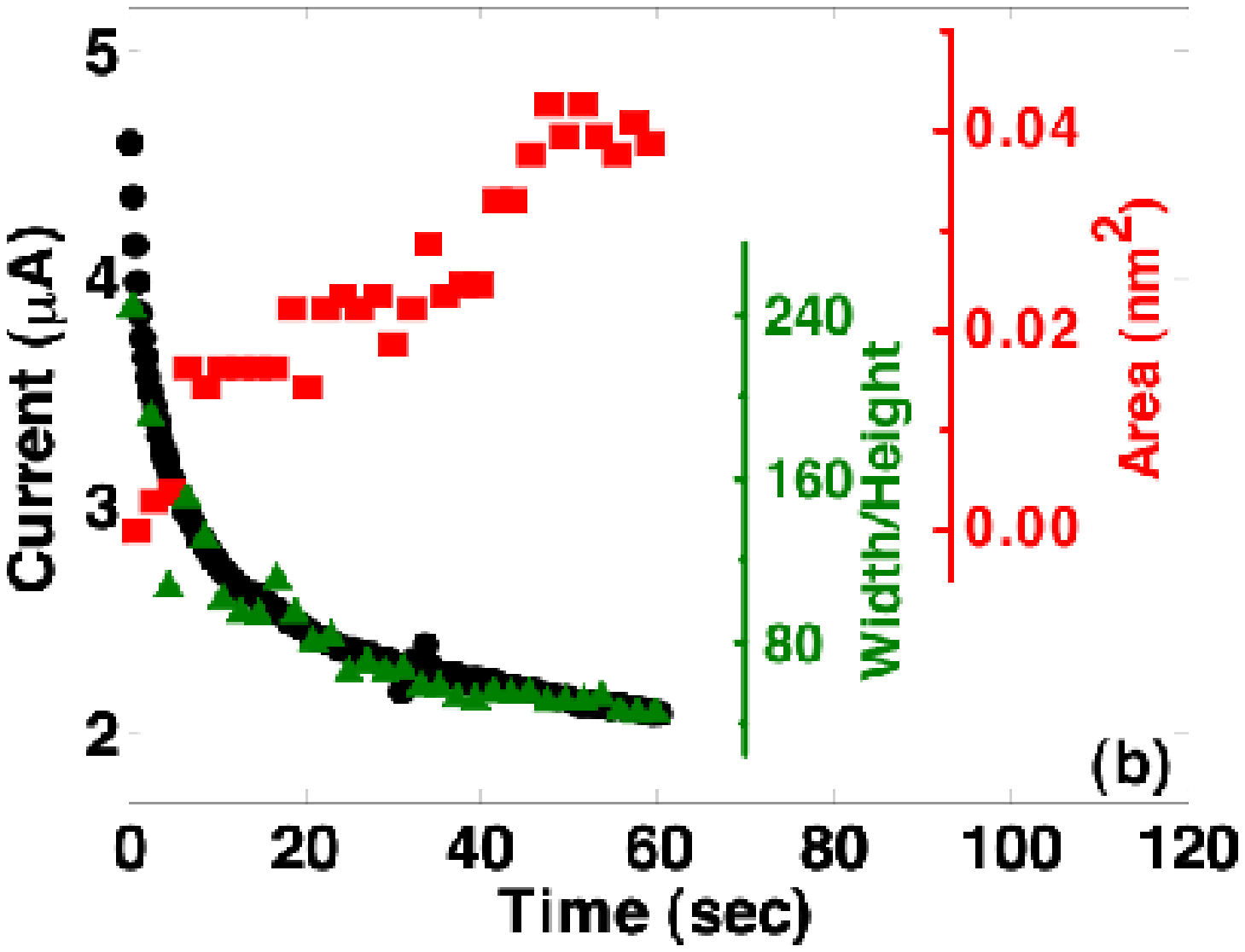}
\vspace{0mm}\\
\caption{ \label{11} (color) In the NaCl solution at $0.25$ V, as
analogous to Figure 5 and 7, (a) the time evolution of a hydrogen
nanobubble is recorded; (b) the current, the nanobubble area, and
the width/height as a function of time are measured. The similar
behaviors as for the pure water case are observed.}
\end{figure*}

\clearpage

\section*{TOC graphic}
\begin{figure*}[h]
\includegraphics[height=65mm]{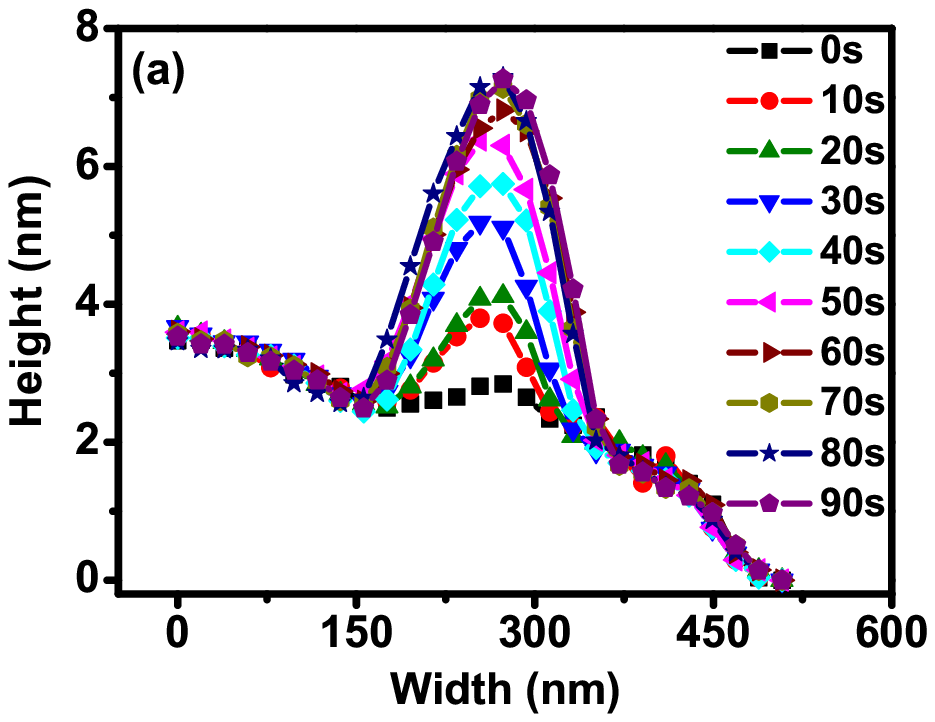}
\includegraphics[height=47mm]{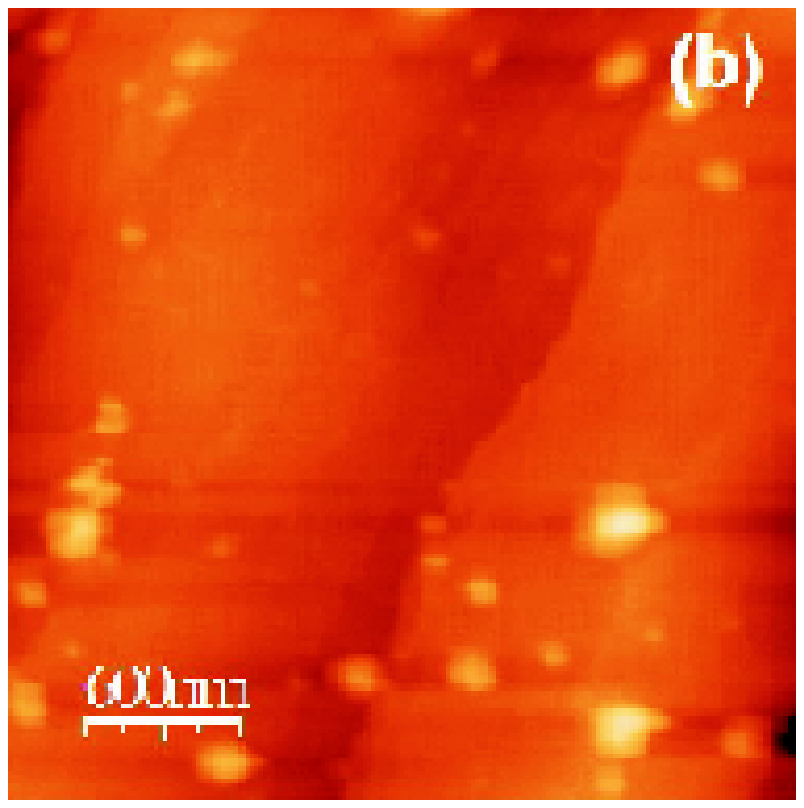}
\vspace{5mm}\\
\end{figure*}

\end{document}